\documentclass[prd,preprintnumbers,nofootinbib]{revtex4}
\usepackage{graphicx}
\usepackage{amsmath}
\usepackage{amsfonts,amsbsy}
\usepackage{amssymb}

\def\lsim{ \,\, \vcenter{\hbox{$\buildrel{\displaystyle <}\over\sim$}}
 \,\,}
\def\be{\begin{equation}}
\def\ee{\end{equation}}
\def\bea{\begin{eqnarray}}
\def\eea{\end{eqnarray}}

\newcommand{\ud}{\, \mathrm{d}}
\newcommand{\xt}{{\mathbf{x}}}
\newcommand{\yt}{{\mathbf{y}}}
\newcommand{\rt}{{\mathbf{r}}}
\newcommand{\bt}{{\mathbf{b}}}
\newcommand{\ut}{{\mathbf{u}}}
\newcommand{\vt}{{\mathbf{v}}}
\newcommand{\kt}{{\mathbf{k}}}
\newcommand{\nabt}{\boldsymbol{\nabla}} 
\newcommand{\tr}{\, \mathrm{tr} \, }

\begin{document}

\title{\bf Anisotropic particle production and azimuthal correlations
  in high-energy pA collisions}

\author{Adrian Dumitru}
\email{Adrian.Dumitru@baruch.cuny.edu}
\affiliation{Department of Natural Sciences, Baruch College, CUNY,
17 Lexington Avenue, New York, NY 10010, USA}
\affiliation{The Graduate School and University Center, The City
  University of New York, 365 Fifth Avenue, New York, NY 10016, USA}

\author{Andre V. Giannini}
\email{avgiannini@usp.br}
\affiliation{Instituto de F\'isica, Universidade de S\~ao Paulo,
C.P.\ 66318, 05315-970 S\~ao Paulo, SP, Brazil\\
Department of Natural Sciences, Baruch College, CUNY,
17 Lexington Avenue, New York, NY 10010, USA}

\author{Vladimir Skokov}
\email{Vladimir.Skokov@wmich.edu}
\affiliation{Department of Physics, Western Michigan University,
  Kalamazoo, MI 49008, USA}

\begin{abstract}
We summarize some recent ideas relating to anisotropic particle
production in high-energy collisions. Anisotropic gluon distributions
lead to anisotropies of the single-particle azimuthal distribution and
hence to disconnected contributions to multi-particle cumulants. When
these dominate, the four-particle elliptic anisotropy $c_2\{4\}$
changes sign. On the other hand, connected diagrams for $m$-particle
cumulants are found to quickly saturate with increasing $m$, a
``coherence'' quite unlike conventional ``non-flow'' contributions
such as decays. Finally, we perform a first exploratory
phenomenological analysis in order to estimate the amplitude ${\cal
  A}$ of the $\cos(2\varphi)$ anisotropy of the gluon distribution at
small $x$, and we provide a qualitative prediction for the elliptic
asymmetry from three-particle correlations, $c_2\{3\}$.
\end{abstract}

\maketitle
\tableofcontents

\section{Introduction}

The observation of large asymmetries, predominantly an elliptic
$\cos(2\varphi)$ asymmetry, in the azimuthal distribution of
particles produced in heavy-ion collisions has been one of the main
indication for the formation of a ``nearly perfect QCD
liquid''~\cite{RHICv2AuAu}. For particles with transverse momenta up
to a few times $\langle p_T\rangle$ this phenomenon is usually
explained in terms of (nearly inviscid) hydrodynamic expansion of an
asymmetric ``fireball''; on the other hand, for high-$p_T$ particles
the asymmetry is thought to originate from energy loss of (mini-) jets
along different paths through the hot and dense Quark-Gluon
plasma~\cite{Shuryak:2004cy}.

More recently, substantial azimuthal asymmetries have also been
observed in p+Pb collisions at the
LHC~\cite{pPb_ALICE,pPb_ALICE2,pPb_ATLAS,pPb_CMS} and in d+Au
collisions at RHIC~\cite{dAu_RHIC}. They are measured via
multi-particle angular correlations (see below) and were found to
extend over a long range in rapidity. By causality, the correlations
must originate from the earliest times of the
collision~\cite{Dumitru:2008wn}. The data shows that the asymmetries
persist up to rather high transverse momenta, well beyond $p_\perp
\sim 1$~GeV. In fact, a recent publication by the ATLAS collaboration
shows that substantial ``elliptic'' ($v_2$) asymmetries in p+Pb
collisions at $\surd s=5$~TeV persist up to $p_\perp =
10$~GeV~\cite{Aad:2014lta}. Final state energy loss is expected
to be much less prominent in smaller systems created in p+p and p+A
collisions; thus it appears reasonable to investigate if azimuthal
asymmetries could originate from the instant of collision when (anti-)
quarks and gluons are ``liberated'' from the wave functions of the
colliding hadrons. Since semi-hard processes involve short-distance
QCD dynamics, we believe that it is important to develop an
understanding of possible origins of azimuthal asymmetries in
perturbative QCD~\cite{Kovchegov:2002nf,pAridge,KovnerLublinsky,Kovchegov:2012nd,Noronha:2014vva,Gyulassy:2014cfa,McLerran:2014uka,Levin:2014kwa,Ozonder:2014sra}.

This paper is a write-up of the talks presented by the authors at the
``Initial Stages 2014'' conference in Napa, CA. It is not a
comprehensive review but attempts to summarize and combine in one paper a few
recent ideas for anisotropic particle production and correlations
within short-distance, small-$x$ QCD.

\section{Scattering of a charge off a semi-classical field}
\label{sec:Scattering}

In the eikonal approximation the S-matrix for scattering of a parton
in the representation ${\cal R}$ of color-SU($N_c$) off the target is
given by~\cite{MuellerDipole}
\begin{equation}
{\cal S}_1(\rt,\bt) \equiv \frac{1}{d_{\cal R}}\tr_{\cal R}\, V^\dagger(\xt)\,
V(\yt)~, \label{Eq:S_1}
\end{equation}
where $\rt \equiv \xt-\yt$ and $\bt \equiv \frac12(\xt+\yt)$ are the
dipole radius and the impact parameter respectively. We have
implicitly assumed that the target field is written in covariant gauge
so that the gauge links from $\xt$ to $\yt$ and back can be dropped.
$d_{\cal R}$ is the dimension of the representation ${\cal R}$ and
$V(\xt)$ denotes a light-like Wilson line describing the propagation
of the parton in the field of the target
\be \label{eq:V_rho}
V(\xt) = \mathbb{P} \exp\left\{ ig \int \ud x^-  
A^{+a}(x^-,\xt) \, t^a_{\cal R}\right\}. 
\ee
Below we shall write most expressions for a fundamental projectile
charge, a quark or anti-quark. The S-matrix for an
adjoint charge (gluon) can be obtained from group theory,
\be
{\cal S}_{\rm A}(\vec r) = \frac{N_c^2 \, |{\cal S}_{\rm F}(\vec r)|^2
  -1}{N_c^2-1} ~.
\ee
While ${\cal S}_{\rm F}(\vec r)$ is complex (for $N_c\ge3$ colors),
${\cal S}_{\rm A}(\vec r)$ is manifestly  real. As we shall see below, this
implies that the single-particle azimuthal distribution for a quark
may in general exhibit odd $v_{2n+1}$ moments while that for a gluon
only has non-zero even moments $v_{2n}$.

Scattering to high transverse momentum corresponds to small
$|\rt|$. This allows us to perform a gradient expansion of the
vector potential $A^+(x^-,\xt)$ resulting in
\be
{\cal S}_1(\rt,\bt) -1 = \frac{(ig)^2}{2N_c}\tr \left( \rt\cdot {\bf
  E}(\bt) \right)^2  
+ \frac{1}{2} \left[ \frac{(ig)^2}{2N_c}\tr \left( \rt\cdot {\bf
    E}(\bt) \right)^2 \right]^2 
+ {\cal O}(r^6)~,
\label{S_1}
\ee
if ${\cal C}$-odd exchanges are dropped. The term of order $r^4$ will
be used in the computation of $c_2\{3\}$ below but is not important
for our main point here. In covariant gauge the light-cone electric
field of the target in Eq.~\eqref{S_1} given by
\begin{equation}
E^i(\bt) = \int dx^- F^{+i} = - \partial^i \int dx^- A^+(x^-, \bt) . 
\label{Eq:E}
\end{equation}

The S-matrix for single parton scattering can be generalized  to  
$m$ particles,
\begin{equation}
{\cal S}_m(\rt_1,\bt_1,\ldots,  \rt_m,\bt_m) -1 =
\left(\frac{(ig)^2}{2N_c} \right)^m \prod_{i=1}^m
 \tr \left( \rt_i\cdot {\bf E}(\bt_i) \right)^2~,
\label{Eq:Sm}
\end{equation}
where we wrote only the leading order in $r$ to simplify the expression.

In the current formalism, event averaging corresponds to averaging
over the target ensemble, which is defined by the field-field
correlator. Conventionally, in the McLerran-Venugopalan
model~\cite{MV} one uses
\begin{equation}
\frac{g^2}{N_c} \langle {E}^a_i(\bt_1) {E}^b_j(\bt_2) \rangle =
\frac{1}{N_c^2-1} \delta^{ab} 
\delta_{ij} \, Q_s^2 \, \Delta(\bt_1-\bt_2)~,
\label{Eq:MVcorr}
\end{equation}
where a general form of the impact parameter dependence of the
correlator $\Delta(\bt)$ with the Fourier image $\tilde{\Delta}(\kt)$
has been introduced. $\Delta(\bt)$ exhibits a logarithmic divergence
as $|\bt|\to0$ which is cut off by the dipole scale $r$ since the
gradient expansion assumes that the electric field is smooth over
scales on the order of the size of the probe.

It should be clear that Eq.~\eqref{Eq:MVcorr} averages over
all fluctuations of the target fields, and hence is isotropic. On the
other hand, for observables which are sensitive to the angular
structure of the target fields, instead we integrate over target field
ensembles subject to the constraint that the anisotropic contribution
to the electric field point in a specific direction $\hat
a$~\cite{KovnerLublinsky,Dumitru:2014dra,Dumitru:2014yza}:
\begin{equation}
\frac{g^2}{N_c} \langle {E}^a_i(\bt_1) {E}^b_j(\bt_2)
\rangle_{\hat{a}} = \frac{1}{N_c^2-1} \delta^{ab} \, Q_s^2\,
\Delta(\bt_1-\bt_2)  \left( \delta_{ij} + 2 {\cal A} \left[\hat{a}_i
\hat{a}_j - \frac{1}{2}\delta_{ij}
      \right] \right)~.
\label{Eq:MVcorrAni}
\end{equation}
That is, we divide the target ensembles into subclasses corresponding
to a particular direction of $\hat{a}$ in the vicinity of the point
with the coordinates $\bt$. The summation over all subclasses
(integration with respect to all possible orientations $\hat{a}$) is
performed {\em after} the observables (such as the $m$-particle
cumulants) have been computed. In other words, the fluctuations from
one configuration ${\bf E}(\xt)$ to another which spontaneously break
2D rotational symmetry constitute {\em slow} variables.

The transverse momentum distribution of scattered partons
can now be written as\footnote{Eq.~(\ref{eq:Def_dN0}) includes the
  ``no scattering'' contribution for transverse momentum exchange
  $k=0$. It plays no role in our subsequent analysis since we are
  interested in finite $k$ only.}
\bea
(2\pi)^2\,
\frac{dN}{k dk \, d\varphi_k}
&=& \int d^2b
  \int d^2r\, e^{-i \vec k \cdot \vec r} \; {\cal
    S}(\rt,\bt)    \label{eq:Def_dN0}\\
&=& \int d^2b \int dr\; r\, d\varphi_r\, e^{-i kr\cos(
\varphi_k-\varphi_r)} \, {\cal S}(r,\varphi_r,\bt) \label{eq:Def_dN1}~.
\eea
The S-matrix satisfies
\be
{\cal S}(r,\varphi_r) = {\cal S}^*(r,\varphi_r+\pi)~.
\ee
Thus, its real part is even under $\varphi_r\to\varphi_r+\pi$
(i.e.\ $\vec{r}\to-\vec{r}$) while its imaginary part is odd.

We can define various asymmetry moments $v_n$ of the single-inclusive
distribution through
\be \label{eq:Def_vn}
v_n(k_T) = \left< \cos n(\varphi_k-\varphi_{\hat a}) \right> = \frac{1}{\cal N}
\int \frac{d\varphi_k}{2\pi} \left<\cos (n(\varphi_k-\varphi_{\hat a}))\, 
\frac{dN}{dy\, k_T dk_T \, d\varphi_k}\right>~,
\ee
with the normalization
\be \label{eq:Norm_kt}
{\cal N} = \int \frac{d\varphi_k}{2\pi} \; \left<\frac{dN}{k_T dk_T \,
  d\varphi_k}\right> = \frac{1}{\pi} 
\left<\frac{dN}{dk_T^2}\right>~.
\ee
The brackets $\langle\cdot\rangle$ indicate an average over all
configurations ${\bf E}(\xt)$.

Even (odd) moments have positive (negative) parity under $\rt\to-\rt$:
\bea
\left< \cos 2n\varphi_k \right> &=& + \left< \cos 2n(\varphi_k+\pi) \right> ~,\\
\left< \cos (2n+1)\varphi_k \right> &=& - \left< \cos (2n+1)(\varphi_k+\pi) \right> ~.
\eea
If ${\cal S}(r,\varphi_r)$ is independent of the orientation of the
dipole then all $v_n=0$. An angular dependence of its real part gives
rise to non-zero parity even moments $v_{2n}$; an angular dependence
of its imaginary part produces odd moments $v_{2n+1}$. For a more
detailed discussion of the $p_T$-dependence of single-particle
$v_1$, $v_2$, $v_3$ we refer to Ref.~\cite{Dumitru:2014dra}.
We note that obtaining non-zero odd-index two-particle cumulants
$v_1\{2\}$, $v_3\{2\}$ as measured in the experiments is more subtle,
see sec.\ \ref{sec:2Pc1c3} below.

\section{Multi-particle cumulants}

\subsection{Connected contributions to high-order cumulants}

We begin this section with the (fully) connected contributions from
$\langle S_m\rangle$ to multi-particle cumulants. We show that these
generate {\em positive} contributions to $c_2\{m\}$ and so would lead
to complex harmonics $v_2\{m\}$ if the number $m$ of particles is a
multiple of four~\cite{Dumitru:2014yza,Skokov:2014tka}. Hence, that
$\langle S_m\rangle$ gives real $v_2\{m\}$ for all $m$ only if the
presence of an azimuthal anisotropy at the single particle level would
generate disconnected contributions.

Furthermore, we show that $|v_2\{m\}|$ beyond $m\simeq4$ is only
weakly dependent on $m$. This indicates a remarkable {\em coherence}
of the connected ``non-flow'' contributions obtained from small-$x$
QCD. Together with their long-range correlation in rapidity, the
properties are quite unlike ``conventional'' non-flow, for example,
from resonance decays or fragmentation of jets.

We then proceed to discuss contributions from fully disconnected diagrams
which arise if rotational symmetry of the single-particle distribution
is broken. These contribute with {\em opposite sign} to $c_2\{2\}$
vs.\ $c_2\{4\}$. We also present a detailed derivation of the
elliptic anisotropy from three-particle correlations, $c_2\{3\}$. This
enables us to analyze a ``BBGKY-like'' hierarchy of $m$-particle
correlations.

The $m$-th order cumulant of the elliptic anisotropy is given by 
\begin{equation}
c_2\{m = 2n\} = \langle \exp\left[i \, 2 (\varphi_1+\varphi_2+\cdots+\varphi_n -
  \varphi_{n+1} - \varphi_{n+2} -  \cdots - \varphi_{2n}) \right] \rangle_\varphi.
\label{Eq:c2m}
\end{equation}
The normalization in Eq.~\eqref{Eq:c2m} is dominated by the
disconnected contributions, corrections are suppressed by powers of $1/N_c^2$.
Thus, after averaging with respect to the impact parameters $\bt_m$ the
normalization at leading order in $N_c$ is
\begin{equation}
\langle S_m (\rt_1, \ldots, \rt_m ) -1  \rangle   \approx\left(
-\frac{Q_s^2}{4} \right)^m \prod_{i=1}^m r_i^2~.   \label{Eq:denominator1}
\end{equation}

Equation~\eqref{Eq:c2m} involves all possible contractions that
generate the fully connected diagrams.  Altogether there are $(2m-2)!!$
contractions:
\begin{eqnarray}
\langle S_m (\rt_1, \bt_1, \ldots, \rt_m, \bt_m) -1 \rangle^{\rm conn.} &=& 
\left(\frac{-Q_s^2}{4}\right)^m
\frac{1}{(N_c^2-1)^{m-1}} 
\Delta(\bt_1-\bt_2) \Delta(\bt_2-\bt_1) \cdots \Delta(\bt_{m-1}-\bt_m)
\Delta(\bt_m-\bt_1) \nonumber \\ && 
(\rt_1 \rt_2) (\rt_2 \rt_3) \cdots (\rt_{m-1} \rt_m) (\rt_m \rt_1)   +
      { [(2m-2)!!-1] \ \rm permutations}.  
\label{Eq:SmConn}
\end{eqnarray}
In what follows we adopt a Gaussian $\Delta(\bt) =
\exp\left( - {\bt^2}/{\sigma^2} \right)$
so that 
\begin{equation}
\frac{1}{S_\perp}\int d^2 b \, \Delta(\bt) = \frac{\pi\sigma^2}{S_\perp}
= \frac{S^c_{\perp} }{S_\perp} = \frac{1}{N_D}~. 
\label{Eq:sigma}
\end{equation}
Here $1/N_D$ is the ratio of the correlated area, $S_\perp^c$, to the
area of the projectile, $S_\perp$ (the proton in p-A collisions),
i.e.\ the inverse number of domains.

Averaging with respect to the impact parameter and angular variables
leads to
\begin{eqnarray}
c_2\{m\} &=& \frac{m!! (m-2)!!}{m \ 2^m} \left[
  \frac{1}{N_D(N_c^2-1)}\right]^{m-1}~~~~,~~~(m\ge2~\mathrm{and~even})~. 
\label{Eq:cFinal}
\end{eqnarray}
The azimuthal harmonics are now readily obtained as~\cite{Skokov:2014tka}:
\begin{equation}
(v_2\{m\})^m  = \frac{(-1)^{\frac{m}{2}+1} }{m\beta_m} \left(
   \frac{1}{N_D(N_c^2-1)} \right)^{m-1}~~~~,~~~(m\ge2~\mathrm{and~even})~,
\label{Eq:Final}
\end{equation}
with
\begin{equation}
 \beta_n = 2 \sum_{k=1}^\infty \left(\frac{2}{j_{0,k}} \right)^{n}~, 
\label{Eq:AssymptBeta}
\end{equation}
where $j_{0,k}$ is the $k$-th zero of Bessel function $J_0(x)$.
Details on transforming the cumulants, $c_2\{m\}$, to the harmonics,
$v_2\{m\}$, can be found in Ref.~\cite{Borghini:2000sa}.
Eq.~\eqref{Eq:Final} also remains true for gluons scattering off the
target owing to the cancelation of Casimir factors in normalized
observables.

\begin{figure}[t]
\begin{center}
\includegraphics[height=7cm]{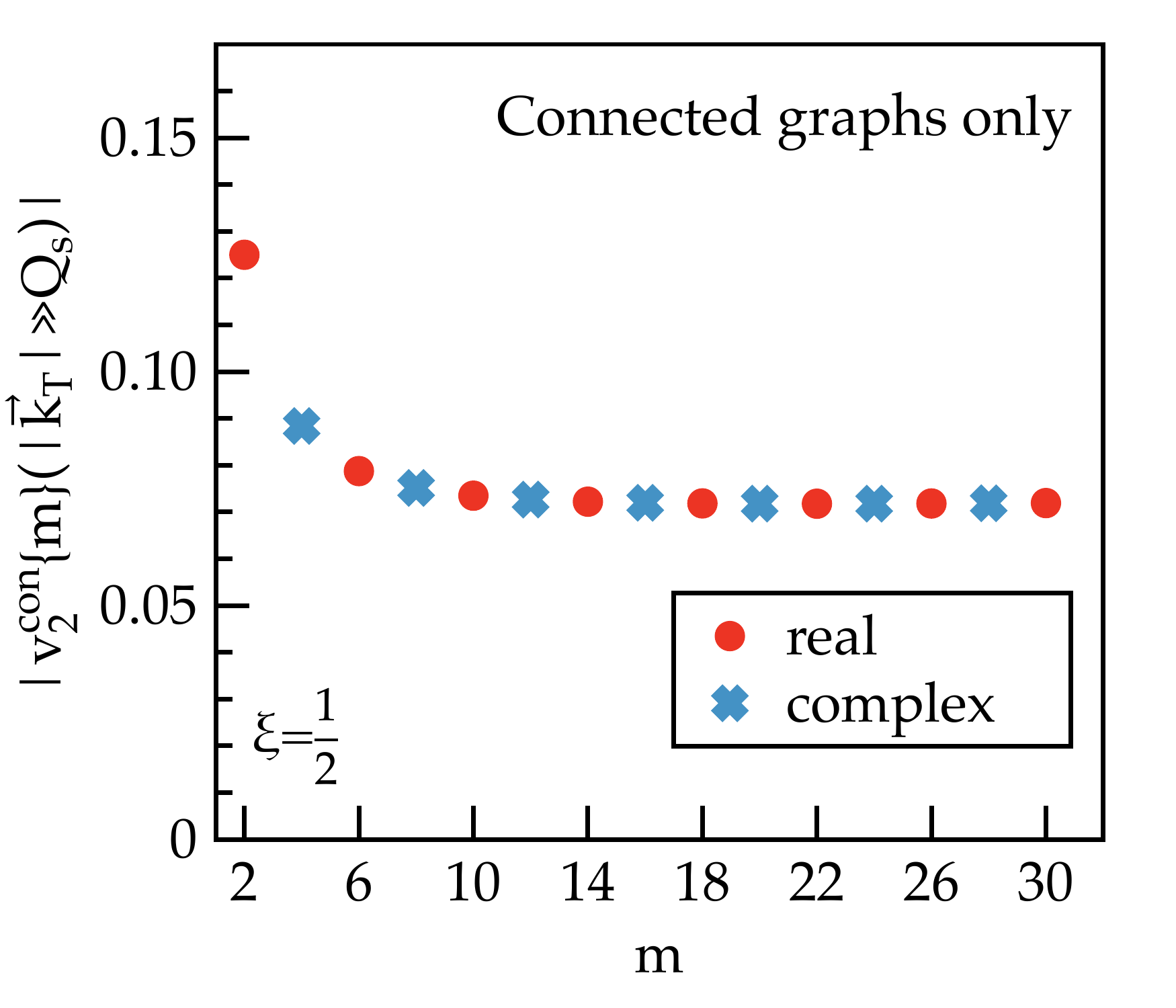}
\includegraphics[height=7cm]{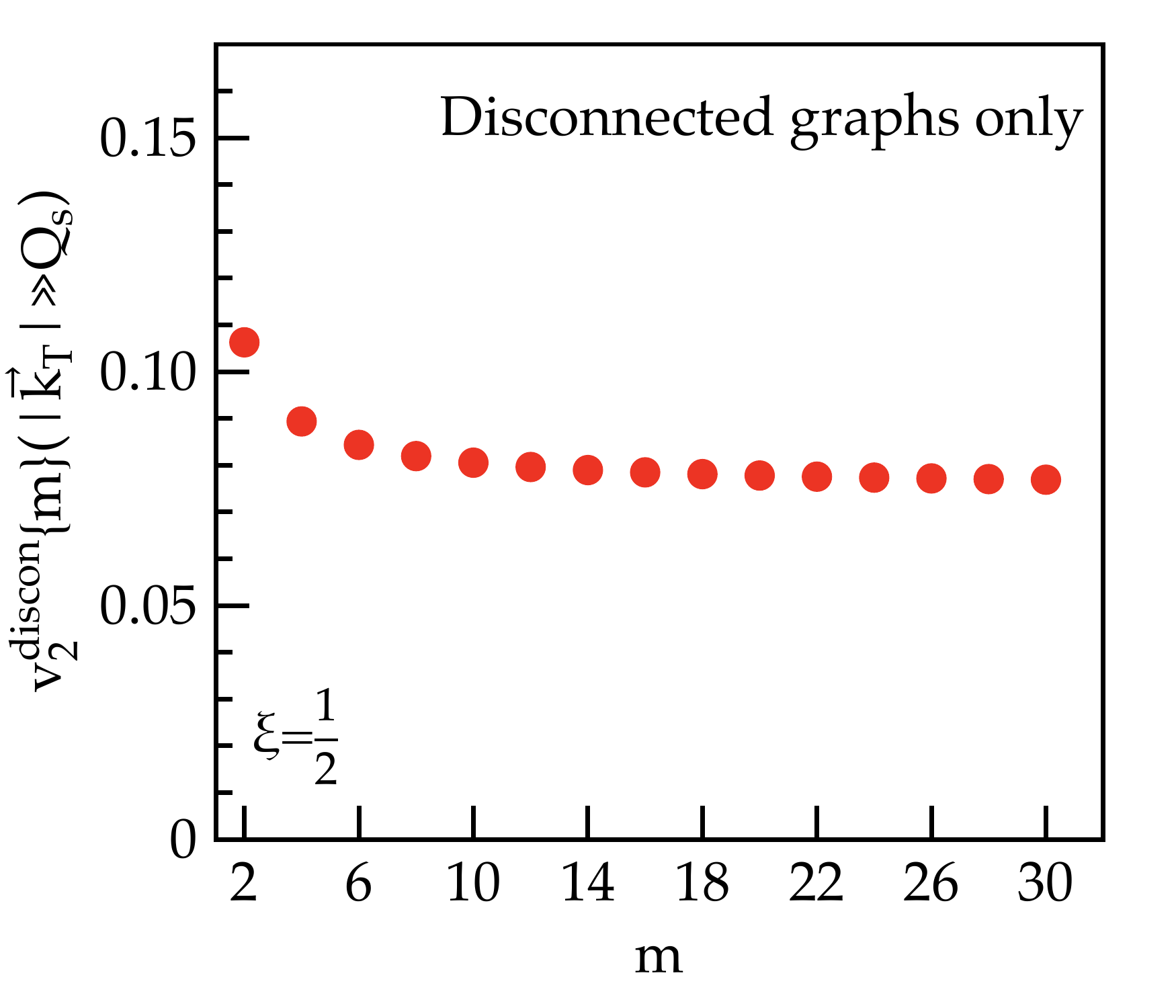}
\end{center}
\vspace*{-0.5cm}
\caption{Left panel: absolute value of $v_2\{m\}$ for connected graphs
  only as a function of $m$. The circles (squares) denote real
  (complex) $v_2\{m\}$.  Right panel: $v_2\{m\}$ for the disconnected
  graphs only, see Eq.~\eqref{Eq:v2m_disc}.  For demonstrational
  purposes the parameter ${\cal A}$ has been chosen such that the
  magnitudes of $v_2^{\rm discon}\{m\}$ and $v_2^{\rm con}\{m\}$ are
  equal at $m\to \infty$. Also, we used $\xi\equiv1/N_D=1/2$.  }
\label{fig:1}
\end{figure}
The absolute values of the harmonics are approximately equal at large
$m$, quickly approaching the limit
\begin{equation}
\lim_{m\to\infty} |v_2\{m\}| = \frac{1}{N_D} \frac{j_{0,1}}{2(N_c^2-1)}~.
\label{Eq:Absolute}
\end{equation}
Although we were unable to prove this rigorously, we believe that this
result holds for any short range correlation $\Delta(\bt)$. The fact
that the disconnected ``non-flow'' small-$x$ diagrams at large $m\ge4$
are of the ``wrong sign'' and approximately independent of $m$ could
perhaps be used to distinguish them from conventional effects.

The other interesting point here is that the fully connected diagrams
give {\it positive} cumulants of {\it any} order and thus, every
second $v_2\{m\}$ is complex, starting from $m=4$:
\begin{equation}
(v_2\{4\})^4 = -  c_2\{4\} = - \frac14
  \left[\frac{1}{N_D(N_c^2-1)} \right]^3 < 0,
\label{Eq:v4}
\end{equation}
This is also illustrated in Fig.~\ref{fig:1}.  A possible resolution
consists in an azimuthal anisotropy of the single dipole
S-matrix~\cite{KovnerLublinsky,Dumitru:2014dra}. This generates
``flow-like'' disconnected contributions to the
cumulants~\cite{Dumitru:2014yza} which we discuss next.

\subsection{Disconnected contributions to the two- and four-particle cumulants}

Equation~(\ref{Eq:MVcorr}) corresponds to averaging over all possible
configurations of $\vec E(\vec b)$ and is isotropic.  However, as we
shall demonstrate in section~\ref{sec:AnisoGluon}, for any particular
configuration the S-matrix does exhibit an angular dependence, see
e.g.\ Fig.~\ref{fig:A_n_b0}. In order to account for this anisotropy
we instead perform the average according to~(\ref{Eq:MVcorrAni}).

The first thing to compute is the angular distribution for scattering
of a single dipole, for fixed $\hat a$. Using the leading term in
Eq.~(\ref{Eq:S_1}) and Eq.~(\ref{Eq:MVcorrAni}), and performing a
Fourier transform to momentum space, as well as an average over the
impact parameter, one arrives at
\be \label{eq:dN1_Fourier_normalized}
\left(\frac{1}{\pi}\frac{dN}{dk^2}\right)^{-1}\; \frac{dN}{d^2k} =
1 - 2{\cal A} + 4{\cal A}\, (\hat k \cdot \hat a)^2~.
\ee
Consequently, the elliptic harmonic of the single-particle
distribution is given by
\be
v_2 \equiv \left< e^{2i(\varphi_k-\varphi_a)}\right>_{\hat a} = {\cal A}~.
\ee
It is straightforward to generalize the computation
of the connected diagrams from above to include the single-particle
anisotropy. The 2- and 4-particle cumulants turn out to be
\begin{eqnarray}
c_2\{2\}&\equiv& (v_2\{2\})^2 
=\frac{1}{N_D}\;
\left({\cal A}^2 + \frac{1}{4(N_c^2-1)}\right)~, 
\label{Eq:c22}
\\ 
c_2\{4\} &\equiv& - (v_2\{4\})^4=  - \frac{1}{N_D^3} \left( {\cal
  A}^4 - \frac{1}{4(N_c^2-1)^3} \right)~. 
\label{Eq:c24}
\end{eqnarray}
The detailed derivation can be found in Ref.~\cite{Dumitru:2014yza}. 

Before presenting the result for the 3-particle cumulant $c_2\{3\}$ we
first examine the results~(\ref{Eq:c22},\ref{Eq:c24}).  The first term
in~\eqref{Eq:c22} is the square of the single-particle $v_2$; it is
scaled by $1/N_D$ since both particles must scatter from the same
domain to exhibit a correlation. The second contribution corresponds
to genuine non-factorizable two-particle correlations, as
discussed above.  Both contributions are positive; nonetheless
Eq.~\eqref{Eq:c22} reveals the existence of two distinct regimes. For
%
${\cal A} \gg \frac{1}{N_c}$
%
the ellipticity is mainly due to the asymmetry of the single-particle
distribution induced by the $\vec E$-field domains. In the opposite limit
%
${\cal A} \ll \frac{1}{N_c}$,
%
$c_2\{2\}$ is mainly due to genuine, non-factorizable two-particle
correlations.

On the other hand, the fourth order cumulant $c_2\{4\}$ changes sign
as a function of ${\cal A}$.  Furthermore, the magnitude of the fully
connected contribution relative to $v_2\{1\}^4$ is $\sim 1/({\cal A}^4
N_c^6)$. Hence, parametrically $c_2\{4\}$ crosses zero when ${\cal
  A}\sim 1/N_c^{3/2}$. Thus, the presence of both connected and
disconnected contributions built from the QCD dipole -- E-field
interaction $\sim \tr (\rt\cdot {\bf E})^2$ can in principle
describe a change of sign of $c_2\{4\}$ as seen in
experiment\footnote{Our result probably does not provide a
  quantitative explanation of the $p_T$-integrated data for $c_2\{4\}$
  which is dominated by particles with low transverse momenta. Also,
  the relation of the anisotropy amplitude ${\cal A}$ and the
  multiplicity is presently not clear.}.

We did not manage to derive the general form of $c_2\{m\}$ for
arbitrary $m$, if both connected and disconnected contributions are
included. However, when the single particle contribution
dominates, one obtains
\begin{equation}
v_2\{m\} = \frac{\cal A} {N_D^{1-1/m}}~.
\label{Eq:v2m_disc}
\end{equation} 
Consequently, in this case, too, the higher order harmonics are
approximately equal to each other, $v_2\{m\} \approx \frac{\cal A}
{N_D}$, for sufficiently large $m$.  We illustrated this in
Fig.~\ref{fig:1} (right).

\subsection{The three-particle cumulant $c_2\{3\}$}
\label{sec:c2_3}

In this section we calculate the quadrupole
anisotropy from 3-particle correlations~\cite{Pruneau:2006gj},
\begin{equation}
v^{3}_{2}\{ 3 \} = c_{2}\{3\} = \langle {\rm exp}\,\,\,
2i(\varphi_{1}+\varphi_{2}-2\varphi_{3})\rangle ~.
\label{v23}
\end{equation}
This cumulant is again defined in such a way as to be invariant under
a simultaneous rotation of all particle transverse momenta by the same
angle.

From Eq.~\eqref{v23} it is clear that the third particle requires a
``$v_{4}$-like" structure or else $v_{2}\{ 3 \}$ would be zero. Such a
contribution $\sim\cos(4\varphi)$ can be obtained from the expansion of
the S-matrix to second order in $\tr (\rt\cdot{\bf E})^2$, see
Eq.~\eqref{S_1}. This leads to the three-dipole S-matrix
\be \label{eq:S3}
\langle S_3 \rangle -1 =
\frac{1}{2}\left(\frac{(ig)^2}{2N_c}\right)^4 \left<
\mathrm{tr}\; (\rt_1\cdot {\bf E}(\bt_1))^2
\; \mathrm{tr}\; (\rt_2\cdot {\bf E}(\bt_2))^2
\; [\mathrm{tr}\; (\rt_3\cdot {\bf E}(\bt_3))^2]^2
\right>~.
\ee
In this case, the most general decomposition of $c_{2}\{3\}$ is given by
\bea
c_{2}\{3\} &=& \langle {\exp}\, (2i (\varphi_1 + \varphi_2 - 2\varphi_3) \rangle^{\rm disc.}
+ \langle {\exp}\, (-4i \varphi_3)\rangle \langle {\exp}\, (2i (\varphi_1 +
\varphi_2) \rangle^{\rm conn.} 
+ 2 \langle \exp (2i\varphi_1)\rangle \langle {\exp}\, (2i (\varphi_2
-2\varphi_3)) \rangle^{\rm conn.} \nonumber\\ 
&+& 2 \langle \exp (2i\varphi_1)\rangle \langle \exp (2i\varphi_3)\rangle
\langle {\exp}\, (2i (\varphi_2 - \varphi_3)) \rangle^{\rm conn.} 
+\langle {\exp}\, (2i (\varphi_1 - \varphi_3)) \rangle^{\rm conn.}\, \langle {\exp}\,
(2i (\varphi_2 - \varphi_3)) \rangle^{\rm conn.} \nonumber \\
&+& \langle {\rm exp}\, (2i (\varphi_1 + \varphi_2 -2\varphi_3)) \rangle^{\rm conn.} \label{c23_out}
\label{c23}
\eea 
Although we have computed all of the above terms here we shall focus
on the fully disconnected contribution $\sim {\cal A}^4$ as well as on
those connected contributions which are of the same order when ${\cal
  A} = \mathcal{O}(N_{c}^{-1})$. The second, third and the last term
in~\eqref{c23} then do not contribute.

The overall normalization implicit in (\ref{c23}) will be approximated
by the angular average of the fully disconnected diagram. It
is given by\footnote{In principle one should Fourier transform first
  to momentum space. At high $p_T$ this transform is simply
  F.T.$\{\rt\} \sim i\kt/k^2$.}
\be
\mathcal{N} = 
-\frac{1}{4^{3}}r_{1}^{2}r_{2}^{2}r_{3}^{2}Q_{s}^{6} ~.
\label{eq:normalization}
\ee
For the fully disconnected contribution we have
\bea
& &
\bigg(\frac{(ig^2)}{2N_c}\bigg)^{4}
\int \frac{d\varphi_{a'}}{2\pi} \int \frac{d\varphi_{a''}}{2\pi}
\left<\mathrm{tr} \left(\vec r_1\cdot \vec E(\vec b_1)\right)^2
\right>_{\hat a}
\left<\mathrm{tr} \left(\vec r_2\cdot \vec E(\vec b_2) \right)^2
\right>_{\hat a'}
\bigg[\left<\mathrm{tr} \left(\vec r_3\cdot \vec E(\vec b_3) \right)^2
\right>_{\hat a''}\bigg]^{2} C(\hat a,\hat a') C(\hat a, \hat a'')\nonumber \\
&=&
\frac{1}{4^{4}}r_{1}^{2}r_{2}^{2}r_{3}^{4}Q_{s}^{8}
\left(1 - {\cal A} + 2{\cal A}\, (\hat r_1 \cdot \hat a)^2\right)
\left(1 - {\cal A} + 2{\cal A}\, (\hat r_2 \cdot \hat a)^2\right)
\left(1 - {\cal A} + 2{\cal A}\, (\hat r_3 \cdot \hat a)^2\right)^{2}
\Delta (\vec{b}_{1}-\vec{b}_{2})\Delta (\vec{b}_{1}-\vec{b}_{3})~,
\label{eq:fully.disc.diagram}
\eea
As in ref.~\cite{Dumitru:2014yza} here we employed $C(a,a') = 2\pi
\delta (a - a') \Delta (\vec{b}_{1}-\vec{b}_{2})$ and $C(a,a'') = 2\pi
\delta (a - a'') \Delta (\vec{b}_{1}-\vec{b}_{3})$ with $\Delta
(\vec{b}_{i}-\vec{b}_{j}) = \exp(-|\vec{b}_{i} -
\vec{b}_{j}|^{2}/\xi^{2})$.  Averaging over the impact parameters
results in
\be
\int \frac{d^2b_1}{S_\perp}\frac{d^2b_2}{S_\perp}\frac{d^2b_3}{S_\perp}
\Delta (\vec{b}_{1}-\vec{b}_{2})\Delta (\vec{b}_{1}-\vec{b}_{3}) =
\frac{\pi \xi^2}{S_{\perp}}\frac{\pi \xi^2}{S_{\perp}} \equiv
\bigg(\frac{1}{N_{D}}\bigg)^2\,,
\ee
with $N_{D}$ the number of ${\bf E}$-field domains in the target
nucleus.

Multiplying~\eqref{eq:fully.disc.diagram} by
$\exp(2i(\varphi_1 + \varphi_2 -2\varphi_3 ))$ and averaging over the azimuthal
angles leads to the disconnected (single particle factorizable) contribution
to $c_{2}\{3\}$; in momentum space,
\be
\langle {\exp}\, (2i (\varphi_1 + \varphi_2 - 2\varphi_3) \rangle^{\rm disc.} =
\frac{1}{N_{D}^{2}}
\frac{1}{4} \frac{Q_{s}^{2}}{k_{3}^{2}} \frac{\mathcal{A}^{4}}{2}~.
\ee
Because we have expanded in the numerator of this cumulant the
S-matrix for the third dipole to second order, we obtain that
$c_{2}\{3\} \sim 1/k_3^2$ drops at high momentum with the square of
the $p_T$ of the third particle.

The connected and disconnected parts of the fourth term in Eq.~\eqref{c23}
are
\bea
&&\frac{(ig)^4}{4 N_c^2}  \left\langle
\mathrm{tr}\;
\left( \vec r_2 \cdot \vec E (\vec b_2)  \right)^2
\; \mathrm{tr}\;
\left( \vec r_3 \cdot \vec E (\vec b_3)  \right)^2
\right\rangle^\mathrm{conn.}_{\hat a}  = \nonumber\\
&& \frac{1}{4^{2}} \frac{r_2^2 r_3^2 Q_s^4}{N_c^2-1} \; \Delta^2(\vec b_2 -\vec b_3)
\left[ \cos(\varphi_2 -  \varphi_3) + 2{\cal A}
  \left( 2\cos\left( \varphi_2 -\varphi_a \right)  \cos\left( \varphi_3 -\varphi_a
  \right) - \cos(\varphi_2-\varphi_3)
  \right)      \right]^2~, \label{eq:2part.conn}\\
&&\frac{(ig)^4}{4 N_c^2}  \left\langle
\mathrm{tr}\;
\left( \vec r_1 \cdot \vec E (\vec b_1)  \right)^2 \right\rangle_{\hat a}
\; \left\langle \mathrm{tr}\;
\left( \vec r_3 \cdot \vec E (\vec b_3)  \right)^2
\right\rangle_{\hat a'}C(a,a') = \nonumber \\
&& \frac{1}{4^{2}}r_{1}^{2}r_{3}^{2}Q_{s}^{4}\,\Delta (\vec{b}_{1}-\vec{b}_{3})
\left(1 - {\cal A} + 2{\cal A}\, (\hat r_1 \cdot \hat a)^2\right)
\left(1 - {\cal A} + 2{\cal A}\, (\hat r_3 \cdot \hat a)^2\right)~, \label{eq:2part.disc}
\eea
respectively. Averaging over impact parameters generates a factor of
$(1/2N_{D}^{2})$. We may now calculate the Fourier transform and sum
over the $4$ contractions of the amplitudes/conjugate amplitudes of
the dipoles $1$ to $3$. This leads to
\bea
&& 2 \langle \exp (2i\varphi_1)\rangle \langle \exp (2i\varphi_3)\rangle
\langle {\exp}\, (2i (\varphi_2 - \varphi_3)) \rangle^{\rm conn.} = 
\frac{1}{N_{D}^{2}}
\frac{1}{4}\frac{Q_{s}^{2}}{k_{3}^{2}}\frac{\mathcal{A}^{2}}{2(N_{c}^{2}-1)} ~.
\eea
The two factors from the fifth term of Eq.~\eqref{c23} each have the
form of Eq.~\eqref{eq:2part.conn}. Averaging over impact parameters,
performing the Fourier transform, and summing over the $8$
contractions of the amplitudes/conjugate amplitudes of the dipoles $1$
to $3$ leads to
\be
\langle {\exp}\, (2i (\varphi_1 - \varphi_3)) \rangle^{\rm conn.}\, \langle {\exp}\,
(2i (\varphi_2 - \varphi_3)) \rangle^{\rm conn.} =
\frac{1}{N_{D}^{2}}\frac{1}{4}\frac{Q_{2}}{k_{3}^{2}}
\frac{1}{16(N_{c}^{2}-1)^{2}}~.
\ee
Finally, for $\mathcal{A}\sim 1/N_{c}$ we have that
\be \label{eq:c23_end}
c_{2}\{ 3 \} = (v_{2}\{3\})^{3} =
\frac{1}{N_{D}^{2}}\frac{1}{4}\frac{Q_{s}^{2}}{k_{3}^{2}}
\bigg( \frac{\mathcal{A}^{4}}{2} + \frac{\mathcal{A}^{2}}{2(N_{c}^2 - 1)} +
\frac{1}{16(N_{c}^{2} - 1)^{2}} \bigg)~.
\ee

As already indicated above, we find that $c_{2}\{ 3 \}\sim 1/k_3^2$ at
high transverse momentum. This is due to the fact that in the
numerator we expanded the S-matrix to order $r_3^4$ while we only
require terms of order $r_3^2$ in the normalization. However,
expression~\eqref{eq:S3} for the S-matrix relies again on the gradient
expansion of the dipole operator. In sec.~\ref{sec:AnisoGluon} below
we shall see that the exact S-matrix (obtained numerically) does
appear to include a $\cos(4\varphi)$ harmonic even at order $r^2$,
indicating the presence of corrections to the gradient expansion.

This provides another way for a ``$v_{4}$-like'' structure at order
$r_3^2$.
In this case $c_{2}\{ 3 \}$ is given by
\bea
c_{2}\{3\} &=& \langle {\exp}\, (2i (\varphi_1 + \varphi_2 - 2\varphi_3) \rangle^{\rm disc.}
+ \langle {\exp}\, (4i \varphi_3)\rangle \langle {\exp}\, (2i (\varphi_1 +
\varphi_2) \rangle^{\rm conn.} 
+ 2 \langle \exp (2i\varphi_1)\rangle \langle {\exp}\, (2i (\varphi_2
-2\varphi_3)) \rangle^{\rm conn.} \nonumber\\ 
&+& \langle {\rm exp}\, (2i (\varphi_1 + \varphi_2 - 2\varphi_3)) \rangle^{\rm conn.} ~.
\label{c23_no_expansion}
\eea
So far we have not yet computed the diagrams involving a contraction
of the third particle with either of the other particles. On the other
hand, it is easy to write down the contributions from the first two terms
in Eq.~\eqref{c23_no_expansion}.

When the third particle is disconnected its S-matrix is given by the
S-matrix for a single dipole and we may decompose its real part into a
Fourier series,
\be
{\cal S}_{1}(\rt_3) - 1 = \mathcal{N}(r_3)\left( 1 +
\sum_{n=1}^{\infty} A_{2n}\cos (2n (\varphi_{r} - \psi))\right)\,.
\ee
The function $\mathcal{N}(r_3) = -r_{3}^{2}Q_{s}^{2}/4$ (at small $r_3$)
is the isotropic part of the S-matrix and $\psi$ is the ``event plane''
angle.

The only term in this series relevant for $c_{2}\{3\}$ is that for $n =
2$. The average over $\langle \exp (-4i\varphi_3)\rangle$ will contribute
with 
\be \langle \exp (-4i\varphi_3)\rangle = A_{4}\int_{-\pi}^{\pi}
\frac{d\varphi_3}{2\pi} e^{-4i(\varphi_3 - \varphi_a)} \cos (4(\varphi_3 -
\varphi_a)) = \frac{A_4}{2}~.
\ee 
The integrand entering the average over the azimuthal angle for the
particles 1 and 2 when they are connected or disconnected has the same
form as in equations \eqref{eq:2part.conn} and \eqref{eq:2part.disc},
respectively. In both diagrams the average over the impact parameters
will generate a factor of $1/N_{D}^{2}$. The overall normalization
factor is given by Eq.~\eqref{eq:normalization} as before.

After computing the averages over the azimuthal angles for all particles
we have that $c_{2}\{3\}$ is now given by
\be
c_{2}\{3\} \sim \frac{1}{N_{D}^{2}} \frac{A_4 \mathcal{A}^{2}}{8}\,.
\label{c23_no_expansion_end}
\ee
We assume that $A_4$ is of order ${\cal A}^2$ and drop
contributions beyond order ${\cal A}^4\sim N_c^{-4}$.
It is evident that if the S-matrix exhibits a
$\cos(4\varphi)$ dependence at order $r^2$ then $c_{2}\{3\}\to$~const
at high transverse momentum. Hence, the behavior of this cumulant at
high $p_{T}$ could provide interesting information about the angular
structure of the dipole S-matrix.

\subsection{BBGKY-like hierarchy of $m$-particle $c_2\{m\}$ cumulants}

In the previous sections we have shown that all $c_2\{m\}$ eventually
are dominated by the fully disconnected contribution proportional to
the single-particle elliptic anisotropy ${\cal A}$ to the $m$-th
power. This occurs in different stages. The four-particle cumulant
$c_2\{4\}$ factorizes when (parametrically) ${\cal A} >
N_c^{-3/2}$. On the other hand, the two-particle cumulant $c_2\{2\}$
requires a stronger $E$-field anisotropy of order ${\cal A} >
N_c^{-1}$. These correlators thus satisfy a BBGKY-like hierarchy. On
the other hand, in the previous section we have seen that the
factorization of $c_2\{3\}$ does not occur at some intermediate value
of ${\cal A}$ but, again, for ${\cal A} > N_c^{-1}$, just like
for $c_2\{2\}$. This correlation function thus represents an exception
to the hierarchy. In sec.~\ref{sec:FitsResults} below we shall attempt
to go beyond parametric estimates of the connected vs.\ disconnected
contributions by performing a phenomenological comparison to data.

\subsection{Odd-index two-particle cumulants, $c_1\{2\}$ and $c_3\{2\}$}
\label{sec:2Pc1c3}

In sec.~\ref{sec:Scattering} we argued that the angular distribution
for a scattered fundamental charge gives rise to odd parity moments $v_1$
and $v_3$. Their dependence on $p_T$ has been discussed and compared
to measured two-particle $v_1\{2\}$, $v_3\{2\}$ in
Ref.~\cite{Dumitru:2014dra}. We would like to point out here that this
issue requires more theoretical investigation, for the following reason.

The two particle correlation function summed over $qq$, $q\bar{q}$,
$\bar{q}q$ and $\bar{q}\bar{q}$ channels is ${\cal C}$-even if one
assumes quark---anti-quark symmetry of the projectile wave function at
small $x$. Indeed the two-particle S-matrix
\begin{equation}
S_2 \propto \left( \tr V^\dagger(\xt_1)\, V(\yt_1) + \tr V(\xt_1)\,
V^\dagger(\yt_1)  \right)
\left( \tr V^\dagger(\xt_2)\, V(\yt_2) + \tr V(\xt_2)\, V^\dagger(\yt_2)  \right)
\label{Eq:S2_real}
\end{equation}
is real, and so has even cumulants only.

Therefore, obtaining non-zero $c_1\{2\}$ and $c_3\{2\}$ may require to
account for (at least) one additional soft rescattering of the (anti-)
quarks besides their hard scattering from the target shockwave. In his
talk at this conference Schlichting showed that classical Yang-Mills
evolution of the liberated gluons in the forward light cone
immediately leads to non-zero $v_3\{2\}$ at time
$\tau=0.1$~fm~\cite{SchlichtingIS2014}. If such rescattering is soft
then the $p_T$-distribution of $v_1$ and $v_3$ shown in
Ref.~\cite{Dumitru:2014dra} should be mostly preserved. Either way,
this clearly is an interesting problem which requires more theoretical
analysis.

\section{Anisotropic gluon distribution at small $x$}
\label{sec:AnisoGluon}

The main goal of this section is to compute scattering of a dipole off
a large nucleus to demonstrate its non-trivial angular
dependence~\cite{Dumitru:2014vka}. We shall first consider the
classical MV model and then proceed to resum quantum fluctuations with
large longitudinal phase space via the JIMWLK evolution equation.

\subsection{Classical McLerran-Venugopalan model}
In the MV model~\cite{MV} the large-$x$ valence partons are viewed as
random, recoilless color charges $\rho^a(\xt)$ described by the
  effective action
\be \label{eq:S2}
S_{\rm eff}[\rho^a] = \int \ud x^- \ud^2{\xt} \; 
\frac{\rho^a(x^-,\xt) \, \rho^a(x^-,\xt)}{2\mu^2}
\ee 
with $\mu^2\sim g^2 A^{1/3}$ proportional to the thickness of a
nucleus; here $A$ denotes the number of nucleons in the nucleus. The
variance of color charge fluctuations determines the average
saturation scale $Q_s^2 \sim g^4 \mu^2$~\cite{JalilianMarian:1996xn}.
The Weizs\"acker-Williams fields generated by $\rho^a(\xt)$ are
pure gauges; in covariant gauge,
\be \label{eq:A+} 
A^{\mu a}(x^-,\xt) = - \delta^{\mu+} \frac{g}{ \nabt^2} \rho^a(x^-,\xt)~.
\ee
Using Eq.~\eqref{eq:A+} in Eqs.~\eqref{Eq:S_1} and \eqref{eq:V_rho} we
can compute the S-matrix for each configuration of the target fields
and extract its Fourier harmonics.

It is rather evident that the random distribution of color charges
$\rho^a(\xt)$ would generate azimuthally anisotropic soft fields. Less
trivially, we shall show that the angular structure of the target
electric fields does not fluctuate randomly on arbitrarily short
scales, i.e.\ that it is characterized by a finite correlation
length $\sim 1/Q_s$ in the transverse plane. This fact is related to
the saturation of the gluon distribution from highly occupied
classical fields at momentum scales $k_T <
Q_s$~\cite{JalilianMarian:1996xn}; over distances $>1/Q_s$ the soft,
classical color fields become ``smooth''. This is why the number of
domains $N_D$ introduced in previous sections is finite.

The most crucial aspect, however, is the following. The angular
structure will obviously fluctuate from one configuration
$\rho^a(\xt)$ of valence charges to the next. Averaging over these
fluctuations like in Eq.~\eqref{Eq:MVcorr} would obviously project
onto the isotropic part of the gluon distribution. Instead, we point
out that the angular fluctuations of $\rho^a(\xt)$ are {\em slow
  variables}, i.e.\ that they should be averaged over only
after the $m$-particle cumulants have been computed.

For a general configuration of the sources, the S-matrix for a
fundamental charge is complex. The real (imaginary) part corresponds
to ${\cal C}$-even (${\cal C}$-odd) interactions~\cite{CGCodderon}:
\bea
1-D_\rho(\rt)\equiv \mathrm{Re}\,{\cal S}_\rho(\rt) =
 \frac{1}{2N_c}\tr \left[V^\dagger(\xt)\,V(\yt) +
   V^\dagger(\yt)\,V(\xt)\right]~,  \label{eq:Def_D}\\
O_\rho(\rt)\equiv  \mathrm{Im}\,{\cal S}_\rho(\rt) =
 \frac{-i}{2N_c}\tr \left[V^\dagger(\xt)\,V(\yt) -
   V^\dagger(\yt)\,V(\xt)\right]~.  \label{eq:Def_iO}
\eea
We use Monte-Carlo techniques on a lattice describing the longitudinal
and transverse coordinates to generate the random configurations
$\rho^a(x^-,\xt)$.  The number of sites in the longitudinal direction
is taken to be $N_- = 100$, while $N_\perp = 1024$ for either of the
transverse directions.  We fix the parameters of the lattice such that
$g^2 \mu a = 0.05$, where $a\equiv L/N_\perp$ denotes the transverse
lattice spacing. Defining the saturation scale from
\begin{equation}
\left<{\cal S}_\rho\right> (r = \sqrt{2}/Q_s) \stackrel{!}{=}
e^{-1/2}
\label{Qsdef}
\end{equation}
we  determined numerically that $Q_s \approx 0.7125 g^2 \mu$.
Further details of the numerical implementation can be found in
Refs.~\cite{Krasnitz:1998ns,Dumitru:2014vka}.

\begin{figure}[t]
\begin{center}
\includegraphics[height=7cm]{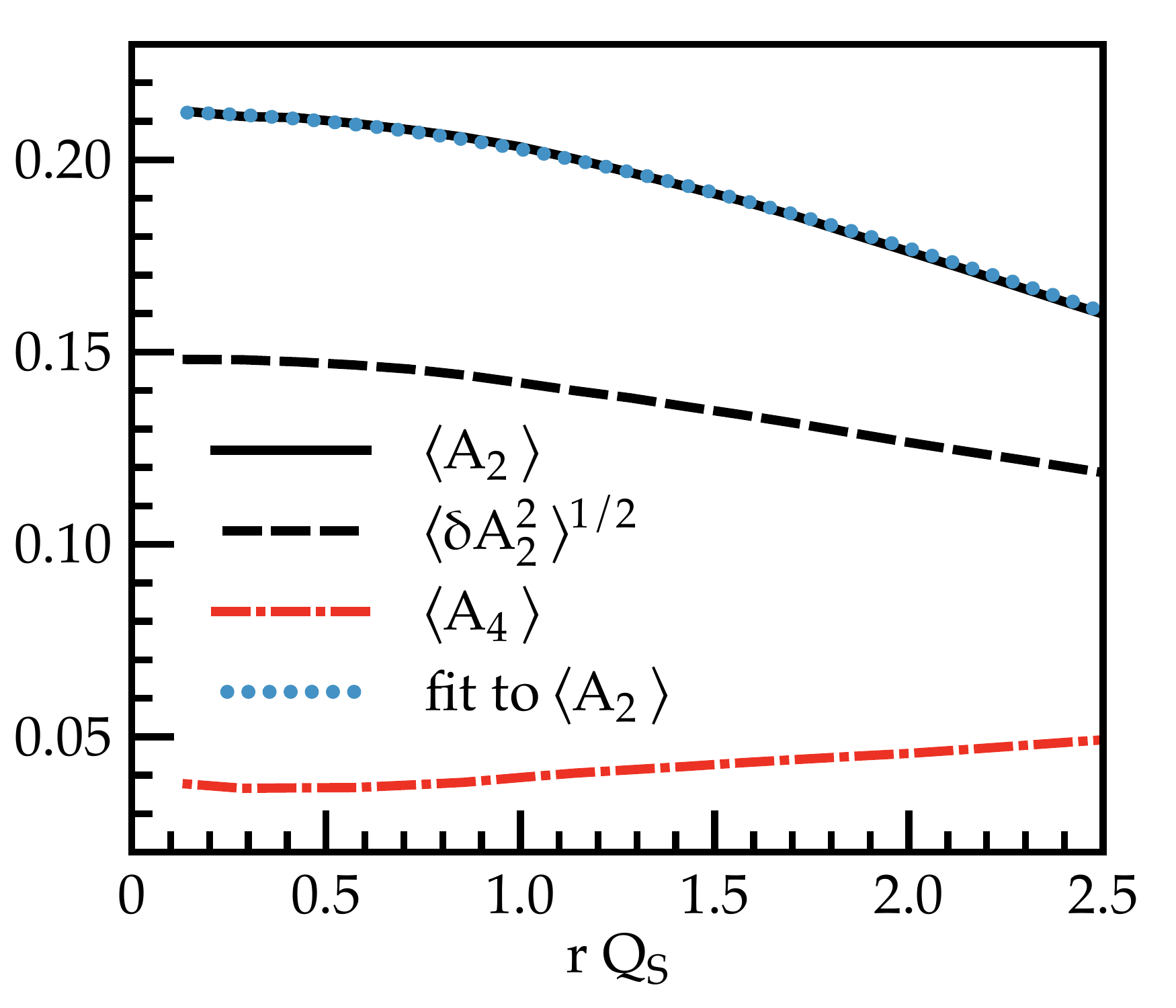}
\includegraphics[height=7cm]{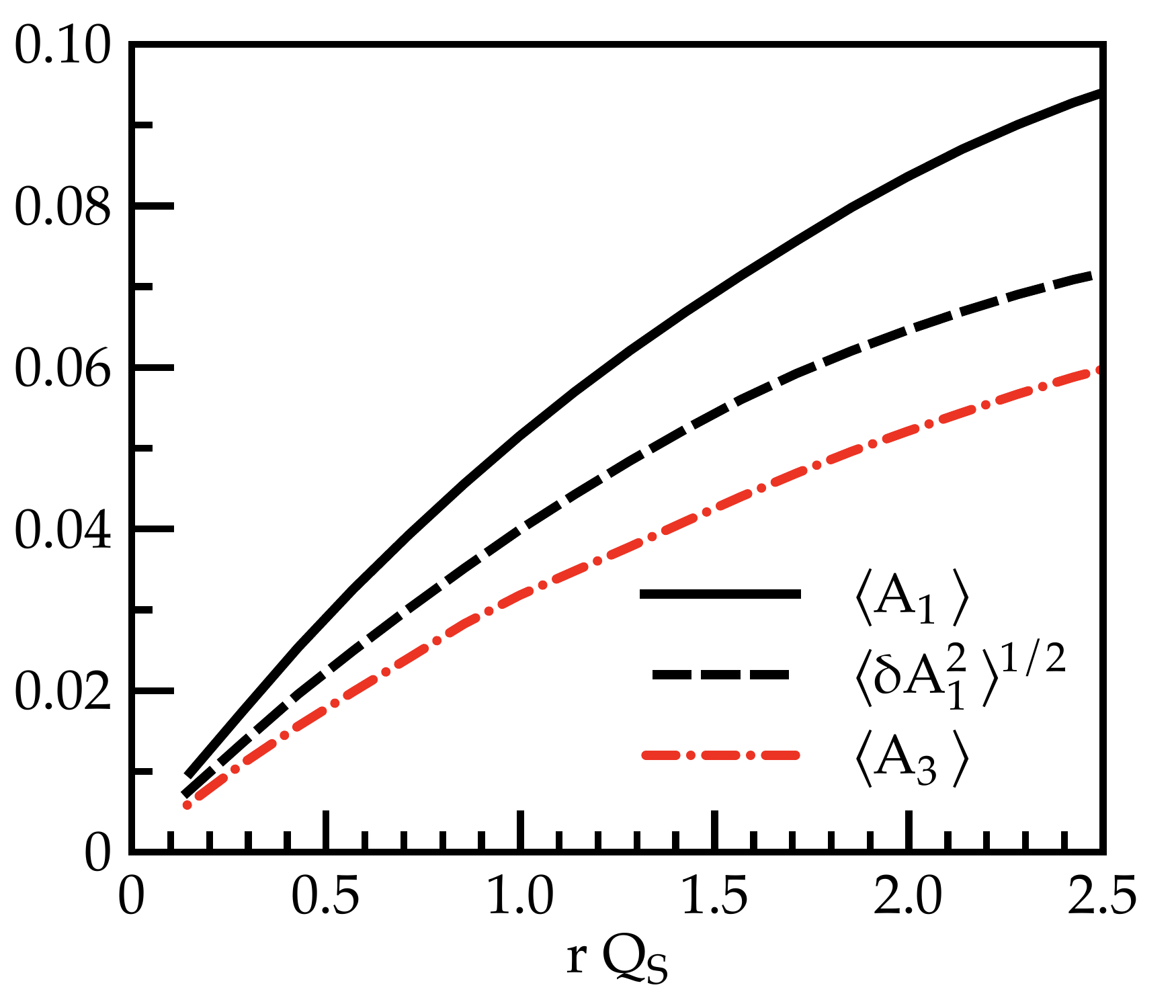}
\end{center}
\vspace*{-0.5cm}
\caption{The configuration-averaged amplitudes $\langle A_n\rangle(r)$
  as functions of the dipole size $r$ for $n=1,\cdots,4$. The fit to
  $\langle A_2\rangle$ corresponds to the function from
  Eq.~(\ref{Eq:h1perp}). Figure from ref.~\cite{Dumitru:2014vka}.}
\label{fig:A_n_b0}
\end{figure}

The azimuthal amplitudes can be extracted by expanding the
real and imaginary parts of the S-matrix in a Fourier series:
\begin{eqnarray}
&& D_\rho(\rt)  = {\cal N}(r)\;  \left( 1  +  \sum_{n=1}^\infty
  A'_{2n} (r) \cos(2 n \varphi_r )  \right)~, \label{eq:Dr_An}\\ 
&&O_\rho(\rt)  = {\cal N}(r)\;  \sum_{n=0}^\infty A'_{2n+1} (r)
  \cos\left[ (2 n+1) \varphi_r \right]~.
\label{Eq:FE}
\end{eqnarray}
Here, the function ${\cal N}(r)$ denotes the isotropic part of the
dipole S-matrix. As already mentioned above each amplitude $A'_n$
contains a random phase which fluctuates from
configuration to configuration. To discard this phase we define
$A_n = \frac{\pi}{2}|A'_n|$;  this arises due to
\be
\int \frac{d\psi}{2\pi}~|\cos \, n\psi| = \frac{2}{\pi}~.
\ee
Then, averaging over $10^4$ configurations we finally obtain $\langle
A_1 \rangle, \cdots, \langle A_4 \rangle$ as well as the variances of
$A_1$ and $A_2$, presented in Fig.~\ref{fig:A_n_b0}.

Our results show that, as expected, the biggest amplitude is the
quadrupole; at $r\lsim 1/Q_s$ the amplitude $\langle A_2\rangle \sim
20\%$.  As we argue in the next section, such values are in the range
of the asymmetries relevant for phenomenology of high-multiplicity
p+Pb collisions at LHC energies.  We stress, however, that in this
calculation we did not attempt to bias the configurations towards
``high multiplicities'', which requires a dedicated investigation. The
function $\langle A_2\rangle(r)$ is almost independent of $r$ for
$r<1/Q_s$ which justifies our treatment in the previous section where
${\cal A}\equiv \langle A_2\rangle$ has been treated as
constant. Figure \ref{fig:A_n_b0} shows furthermore that the variance
$\surd \langle (\delta A_2)^2\rangle$ is similar in magnitude to the
mean value $\langle A_2\rangle$. This points at rather large
fluctuations of $A_2$ for different configurations.

\begin{figure}[htb]
\begin{center}
\includegraphics[height=7cm]{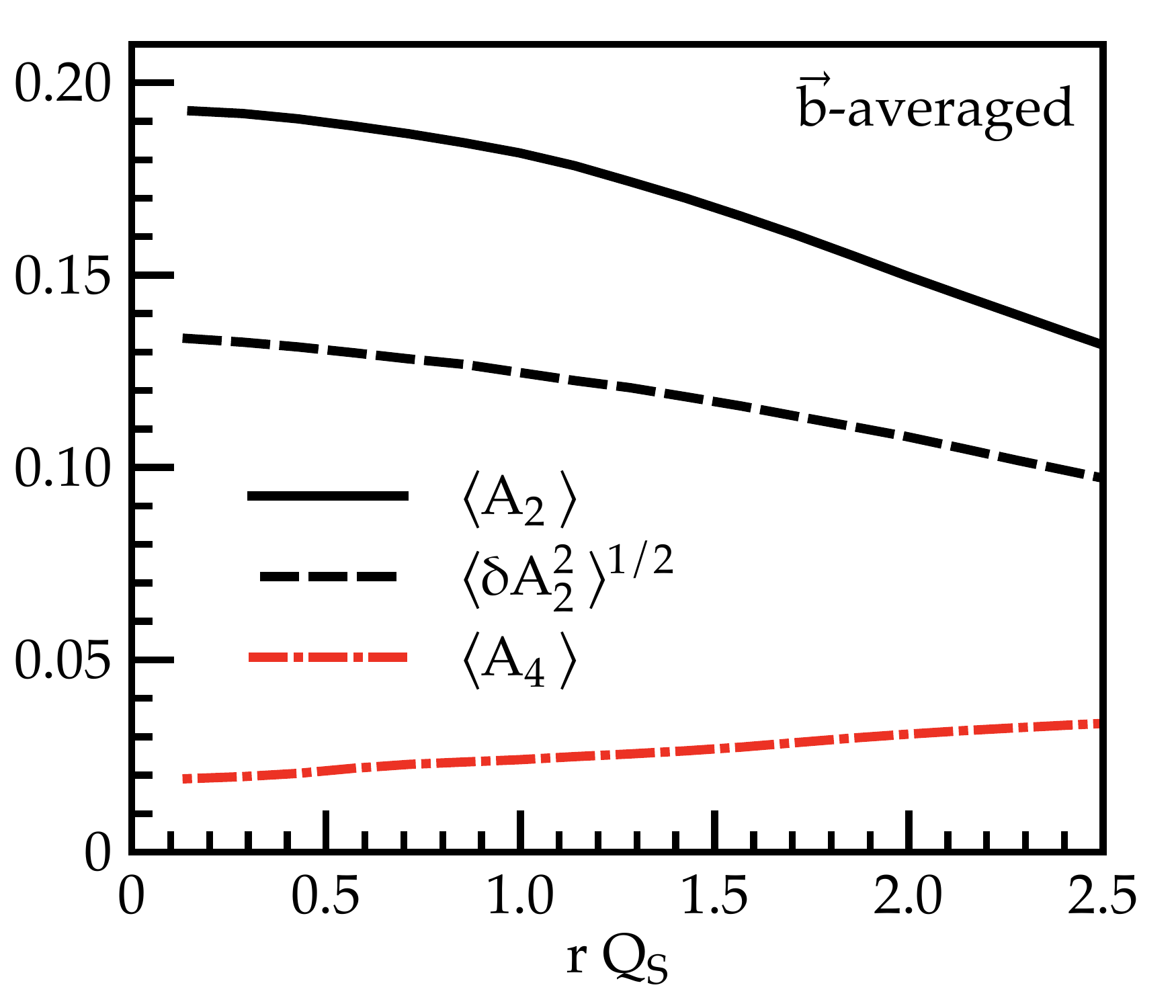}
\includegraphics[height=7cm]{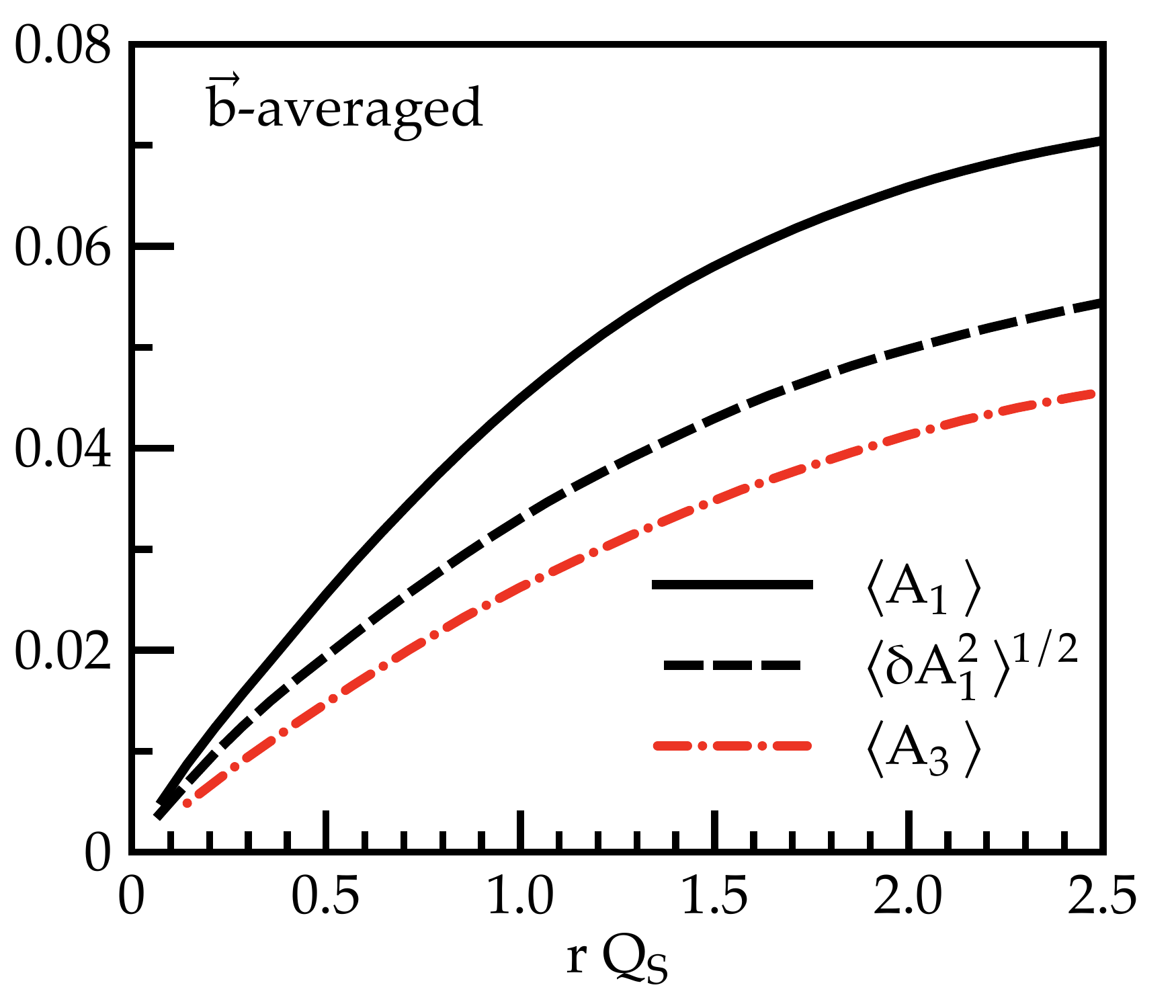}
\end{center}
\vspace*{-0.5cm}
\caption{Same as Fig.~\ref{fig:A_n_b0} for ``$\bt$-smeared'' target
  ${\bf E}$-fields.}
\label{fig:A_n_b}
\end{figure}
Figure~\ref{fig:A_n_b} shows the same amplitudes as the previous figure
but for ${\bf E}$-fields which have been ``smeared'' over an area $\pi
r^2$ set by the size of the dipole. Comparing Figs.~\ref{fig:A_n_b0}
and~\ref{fig:A_n_b} one sees that ``smearing'' has a negligible effect
for $r\lsim 1/Q_s$ while the anisotropy amplitudes at large $r$ are
suppressed. This behavior shows the correlation over finite transverse
distance scales of the angular structure of the ${\bf E}(\xt)$
configurations.

Reference~\cite{Dumitru:2014vka} showed that the MV-model amplitude
$\langle A_2\rangle(r)$ matches the distribution of linearly polarized
gluons (for an unpolarized target) $h_1^{\perp g}(x,\kt^2)$ introduced
in TMD factorization~\cite{TMD,TMD2} 
\be 
\delta^{ij}f_1^g(x,\kt^2) +
\left(\hat{k}^i\hat{k}^j-\frac{1}{2}\delta^{ij}\right) h_1^{\perp
  g}(x,\kt^2)~.  
\ee 
Within the framework of the MV model, the result for $h_1^{\perp g}(x,
r)$ derived analytically in Ref.~\cite{TMD2},
\begin{equation}
h_1^{\perp g}(x, \rt^2) \propto \frac{1}{r^2 Q_s^2} \left[ 1 -
  \exp\left( - \frac{r^2 Q_s^2}{4}  \right) \right]~,
\label{Eq:h1perp}
\end{equation}
is in good agreement with our numerical results at small values
of $r\lsim 2  Q_s^{-1}$.
  
Figure~\ref{fig:A_n_b0} also shows a non-zero amplitude of the $\cos
(4\varphi)$ angular component. It appears to be essentially constant at
small $r$ unlike the $\sim r^2$ behavior expected from the second term
in Eq.~\eqref{S_1} once scaled by ${\cal N}(r)\sim (rQ_s)^2$ at small
$r$. This may be due to corrections to the gradient expansion which
was used to derive Eq.~\eqref{S_1}. Such a term would provide another
contribution to the hexadecupole $v_4$-like asymmetry, c.f.\ previous
section.

Due to fluctuations of the saturation momentum $Q_s$ in impact
parameter space~\cite{Kovchegov:2012ga} every particular configuration
of semi-classical small-$x$ fields~(\ref{eq:A+}) contains a ${\cal
  C}$-odd component and $O(\rt)$ as defined in Eq.~(\ref{eq:Def_iO})
is non-zero. This results in non-zero odd-index amplitudes $A_1$ and
$A_3$, see Fig.~\ref{fig:A_n_b0}. The figure also shows that
the expectation values of the odd amplitudes are significantly smaller
than $A_2$; as expected, they vanish as $r\to0$:
\be
iO(\rt) \sim i\, \alpha_s\, \rt\cdot\nabt_\bt \left(
1 - D(\rt,\bt)\right)
\simeq i\, \alpha_s\, r^3\, Q_s^2 \, Q_c\, \cos \varphi_r \left[
1 - \frac{r^2}{4} \left( \frac{Q_c^2\cos^2 \varphi_r}{3} + Q_s^2
  \right)
\right]~.
\label{eq:iO_r3}
\ee
The expression on the r.h.s.\ corresponds again to a gradient
expansion in powers of $r$, assuming a generic spectrum of
fluctuations of $Q_s(\bt)$ cut off at $Q_c$~\cite{Dumitru:2014dra}.

The presence of odd harmonics does not indicate that the expectation
value of the ${\cal C}$-odd part of the S-matrix is non-zero. Indeed,
the average of the odderon $O(\rt)$ over the ${\cal C}$-even ensemble
generated by the action~(\ref{eq:S2}) is zero. However, the product of
$O(\rt)$ with another ${\cal C}$-odd operator, which effectively
arises due to our dropping of the phases of the amplitudes $A_n'$, is
even under ${\cal C}$-conjugation and its expectation value is not
zero.

\subsection{Quantum fluctuations and high-energy evolution}

In the previous subsection, within the framework of the classical MV
model, we showed that scattering of a dipole from the soft fields
sourced by a particular configuration $\rho^a(\xt)$ of valence charges
is not isotropic, and that the amplitudes of the azimuthal anisotropy
are quite significant.  In this section we consider how these
amplitudes are affected by small-$x$ / high energy evolution.
This corresponds to a resummation of (nearly) boost-invariant quantum
fluctuations to the classical field.

The evolution of the elliptic anisotropy with rapidity was first
addressed by Kovner and Lublinsky in Ref.~\cite{KovnerLublinsky}. They
solved the BK evolution equation for the dipole scattering amplitude
$1-{\cal S}(\rt)$ as a function of dipole size {\em and}
orientation. They found that the anisotropy decays exponentially with
$Y=\ln(x_0/x)$. Their solution, however, was based on the assumption
that the impact parameter space is homogeneous. As explained in the
previous section, even at the level of the initial condition (given by
the MV model), the azimuthal anisotropy of ${\cal S}(\rt,\bt)$ arises due
to fluctuations of the soft fields in the transverse impact
parameter plane. Hence, in this subsection we describe solutions of
JIMWLK evolution which account for fluctuations of the light-like
electric Wilson lines in $\bt$-space~\cite{Dumitru:2014vka}.

Going beyond the classical theory, quantum gluon emissions which are
enhanced by a large longitudinal phase space $Y=\log x_0/x$ are
resummed by the so-called JIMWLK~\cite{jimwlk,JIMWLK_rm1} functional
renormalization group evolution. It modifies the ensemble of electric
Wilson lines over which observables are averaged thereby resumming
corrections to all orders in $\alpha_s Y$. Evolution over a step
$\Delta Y$ in rapidity opens up phase space for radiation of
gluons and modifies the classical action~(\ref{eq:S2}). The
evolution can be formulated in terms of a ``random walk'' in the space
of Wilson lines
$V(\xt)$~\cite{JIMWLK_rm1,Lappi:2012vw}:
\bea
\partial_Y V(\xt) =
V(\xt)
\frac{i}{\pi}\int \ud^2\ut
 \frac{(\xt-\ut)^i\eta^i(\ut)}{(\xt-\ut)^2}
- \frac{i}{\pi}\int \ud^2\vt 
  V(\vt) \frac{(\xt-\vt)^i \eta^i(\vt)}{(\xt-\vt)^2} V^\dag(\vt) 
V(\xt)~.
\label{eq:Lgvn}
\eea
The Gaussian white noise $\eta^i =\eta^i_a t^a$ satisfies
$\langle \eta^a_i(\xt)\rangle =0$ and
\be \label{eq:etaeta}
\langle \eta^a_i(\xt)\; \eta^b_j(\yt)\rangle = \alpha_s \, \delta^{ab}
\delta_{ij}\delta^{(2)}(\xt-\yt).
\ee
The so-called ``left-right symmetric'' form of Eq.~(\ref{eq:Lgvn}) was
introduced in Ref.~\cite{Kovner:2005jc}. We solve
Eq.~\eqref{eq:Lgvn} numerically assuming a fixed but small coupling
$\alpha_s=0.1$; for such coupling the speed of evolution is at least
roughly comparable to more realistic running coupling evolution. 

Once an ensemble of Wilson lines on the transverse lattice has been
evolved to rapidity $Y$, we can again compute the dipole scattering
amplitude ${\cal S}_Y(\rt,\bt)$, its azimuthal Fourier decomposition
and the corresponding saturation scale $Q_s(Y)$ using
Eq.~\eqref{Qsdef}.  It is important to note here that even though we
consider a target of infinite transverse extent (periodic boundary
conditions), that the evolution equation is solved on a transverse
lattice which does allow for impact parameter dependent fluctuations.
\begin{figure}[htb]
\begin{center}
\includegraphics[height=7cm]{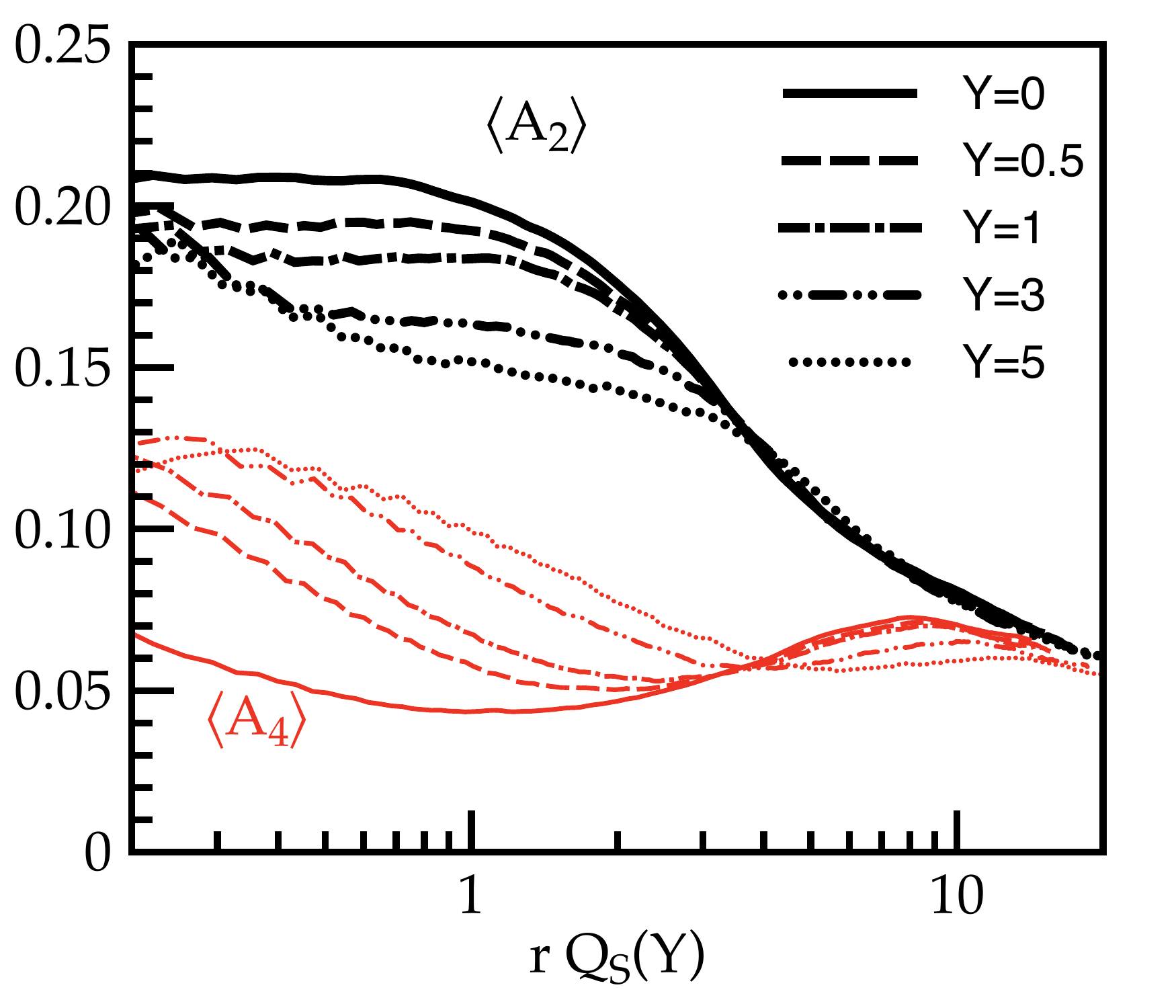}
\includegraphics[height=7cm]{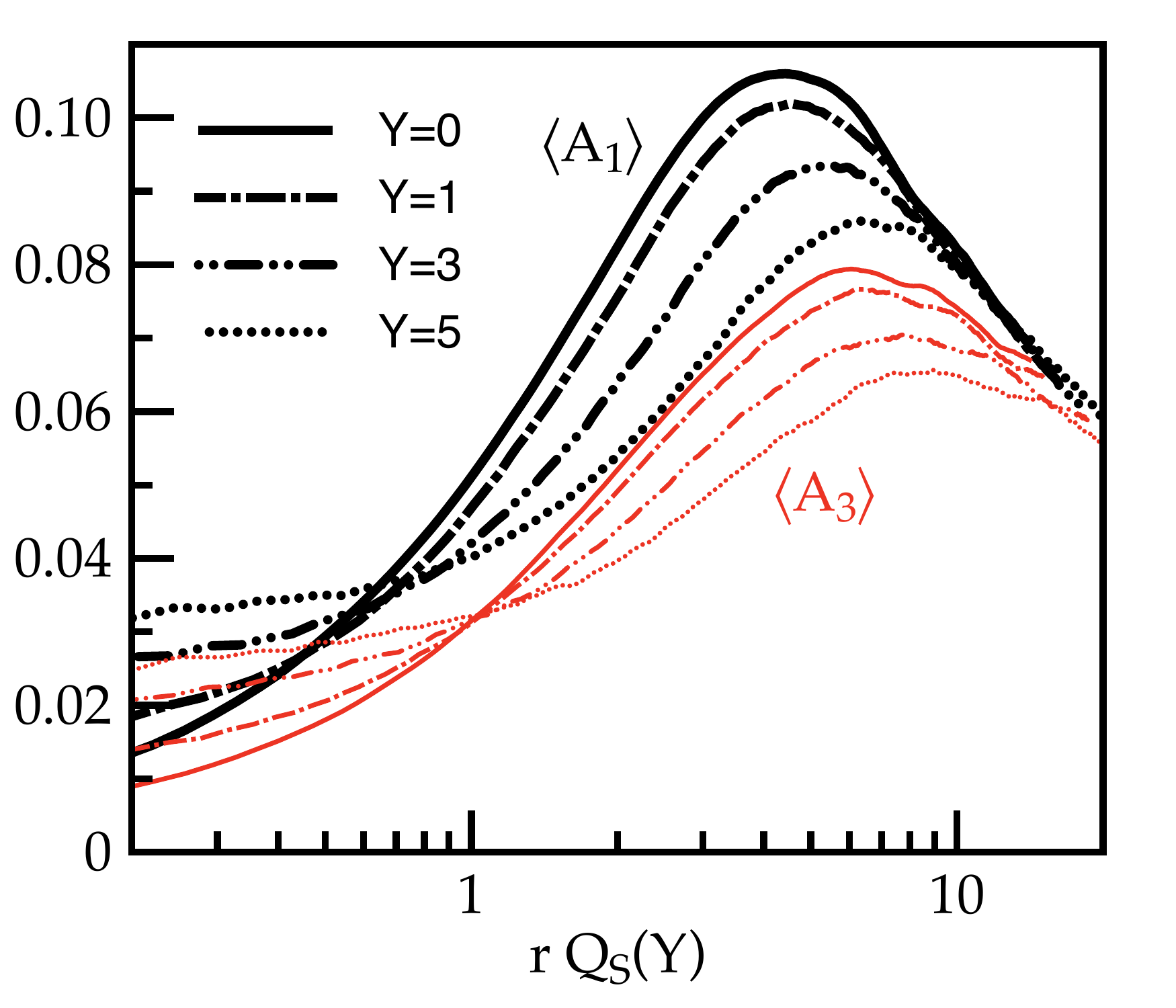}
\end{center}
\vspace*{-0.5cm}
\caption{JIMWLK evolution of $\langle A_2\rangle(r)$ and $\langle
  A_4\rangle(r)$ (left) resp.\ of $\langle A_1\rangle(r)$ and $\langle
  A_3\rangle(r)$ (right). The lower order harmonics
  correspond to the upper sets of curves. Figure from
  Ref.~\cite{Dumitru:2014vka}.}
\label{fig:A_n_jimwlk}
\end{figure}

\begin{figure}[htb]
\begin{center}
\includegraphics[height=7cm]{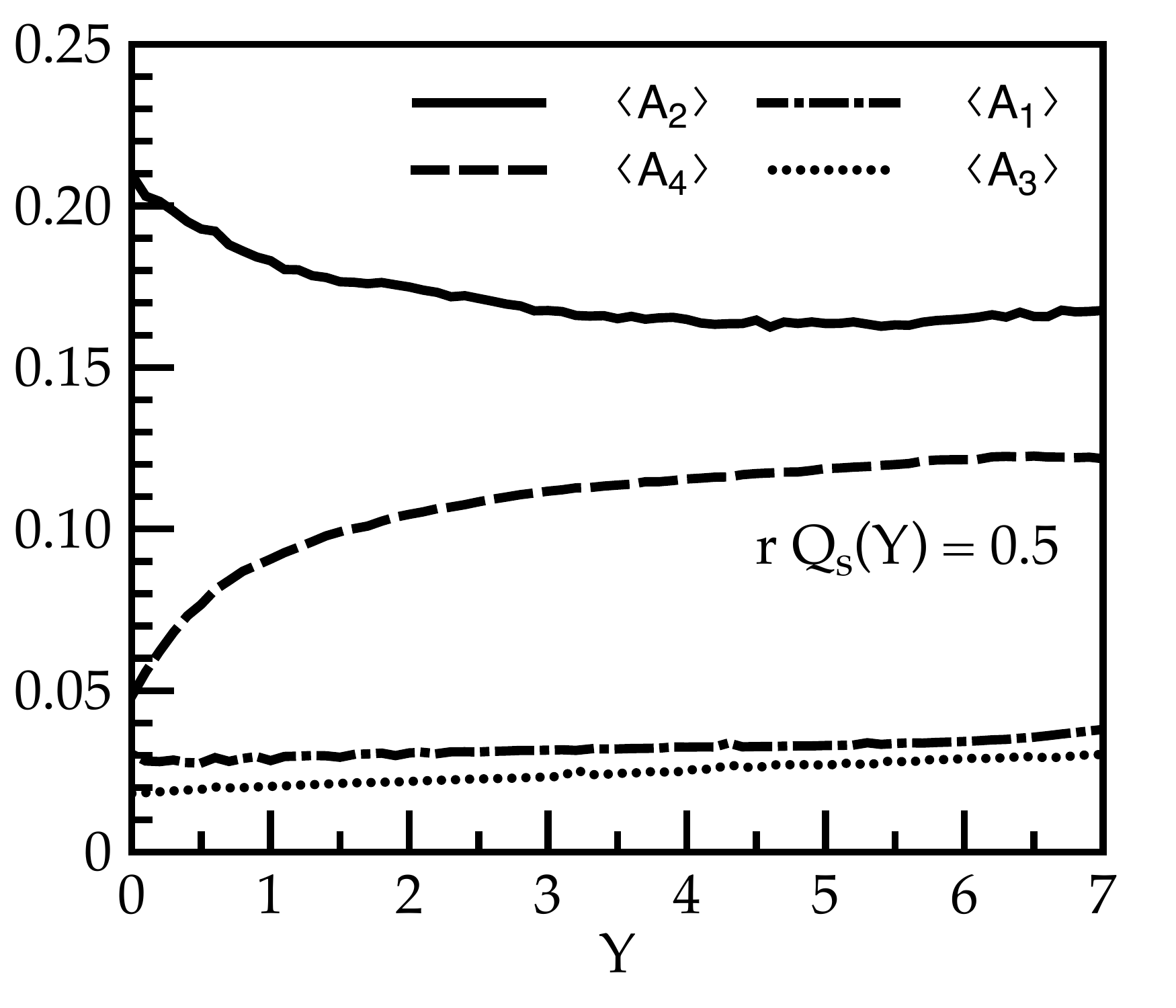}
\includegraphics[height=7cm]{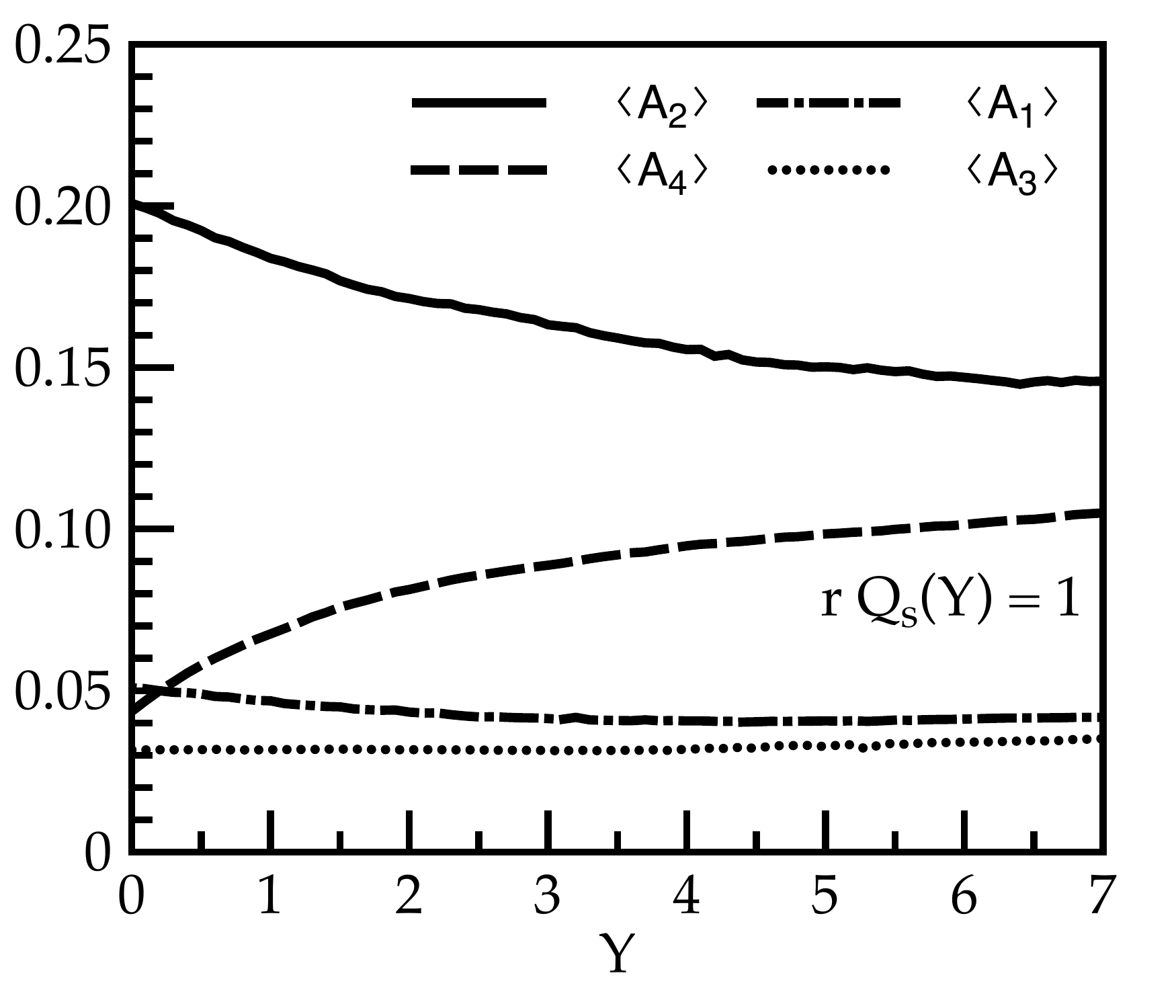}
\end{center}
\vspace*{-0.5cm}
\caption{JIMWLK evolution of $\langle A_{1,2,3,4}\rangle$
at fixed $r Q_s(Y)$. }
\label{fig:A_Y_jimwlk}
\end{figure}
In Fig.~\ref{fig:A_n_jimwlk} (left) we show the evolution of $\langle
A_2\rangle(r)$ and $\langle A_4\rangle(r)$ with $Y$. As already
mentioned above, mean-field evolution of the dipole was shown to wash
out initial elliptic anisotropies rather
quickly~\cite{KovnerLublinsky}. On the other hand, here we only
observe a relatively slow decrease of $\langle A_2\rangle(r)$ with
$Y$. This is rather intuitive since both the initial anisotropies at
$Y=0$, as well as those of the evolved JIMWLK configurations are
generated by fluctuations of the hard ``valence charges'' in the
transverse impact parameter plane. Furthermore, we observe that those
harmonics which are initially small, i.e.\ $\langle A_1\rangle(r)$,
$\langle A_3\rangle(r)$ and $\langle A_4\rangle(r)$, in fact increase
with $Y$ at small $r$. The harmonics also display universal behavior
at very large $r$. The evolution of the amplitudes with $Y$ at fixed
$r\, Q_s(Y)$ is shown in Fig.~\ref{fig:A_Y_jimwlk}.

Thus, we conclude that the anisotropies are not washed out by high
energy evolution and that they might be essential to describe short
distance azimuthal asymmetries observed at LHC. An initial
phenomenological analysis is presented in the next section.

\section{Application to phenomenology of proton-nucleus  collisions}
\label{sec:FitsResults}

In this section we present a first phenomenological comparison of
the measured $v_2\{2\}$ and $v_2\{4\}$ at high transverse momentum to
some of the expectations from above. Our analysis is certainly not
definitive but preliminary and qualitative. Our main goals are:
\begin{itemize}
\item to check if the magnitudes of $v_2\{2\}$ and $v_2\{4\}$ can be
  reproduced for ``reasonable'' values of $N_D$, the number of ${\bf
    E}$-field domains, and of ${\cal A}$, the ${\bf E}$-field
  $\cos(2\varphi)$ anisotropy amplitude;
\item to verify that the connected contributions to the two- and
  four-particle cumulants indeed describe the splitting between
  $v_2\{2\}$ and $v_2\{4\}$ observed experimentally at semi-hard
  $p_T$;
\item to estimate the relative magnitudes of connected
  vs.\ disconnected contributions to the two-, three-, and
  four-particle cumulants, i.e.\ how far the respective cumulants are
  from the factorization limit (dominance of fully disconnected diagrams);
\item to make a prediction for $v_2\{3\}$ in pA collisions at the LHC.
\end{itemize}

In order to fix $\mathcal{A}$ and $N_{D}$, we shall use the CMS
$v_{2}(p_{T})$ data from $2$- and $4$-particle correlations in p+Pb
collision at 5~TeV. We focus on the highest multiplicity
events. Equations~\eqref{Eq:c22} and~\eqref{Eq:c24} provide the
theoretical expectations for the cumulants $c_{2}\{2\}$ and
$c_{2}\{4\}$ which we repeat here for convenience:
\begin{eqnarray}
c_2\{2\}&\equiv& (v_2\{2\})^2 
=\frac{1}{N_D}\;
\left({\cal A}^2 + \frac{1}{4(N_c^2-1)}\right)~, 
\label{Eq:c22b} \\ 
c_2\{4\} &\equiv& - (v_2\{4\})^4=  - \frac{1}{N_D^3} \left( {\cal
  A}^4 - \frac{1}{4(N_c^2-1)^3} \right)~. 
\label{Eq:c24b}
\end{eqnarray}
\begin{figure}[htb]
\begin{center}
\includegraphics[width=8cm]{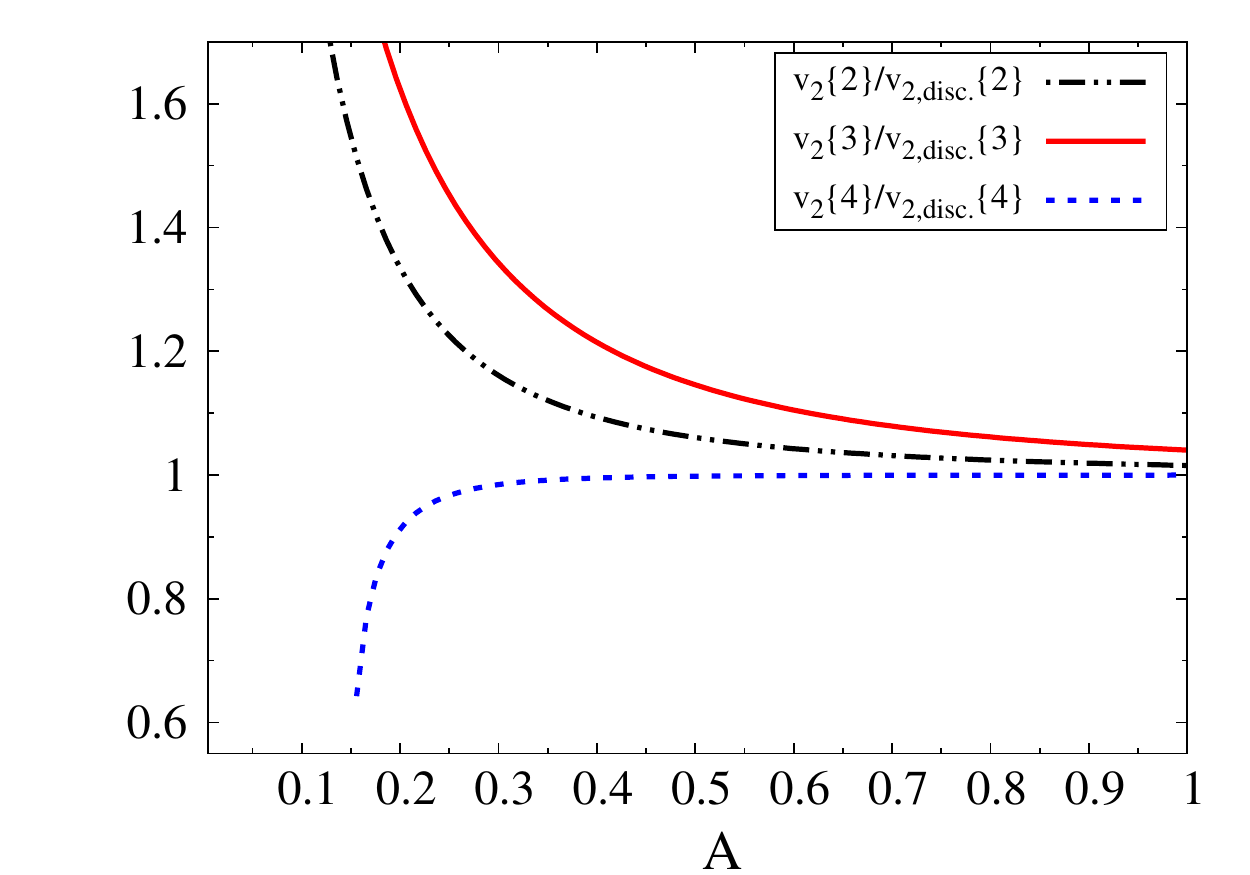}
\includegraphics[width=8cm]{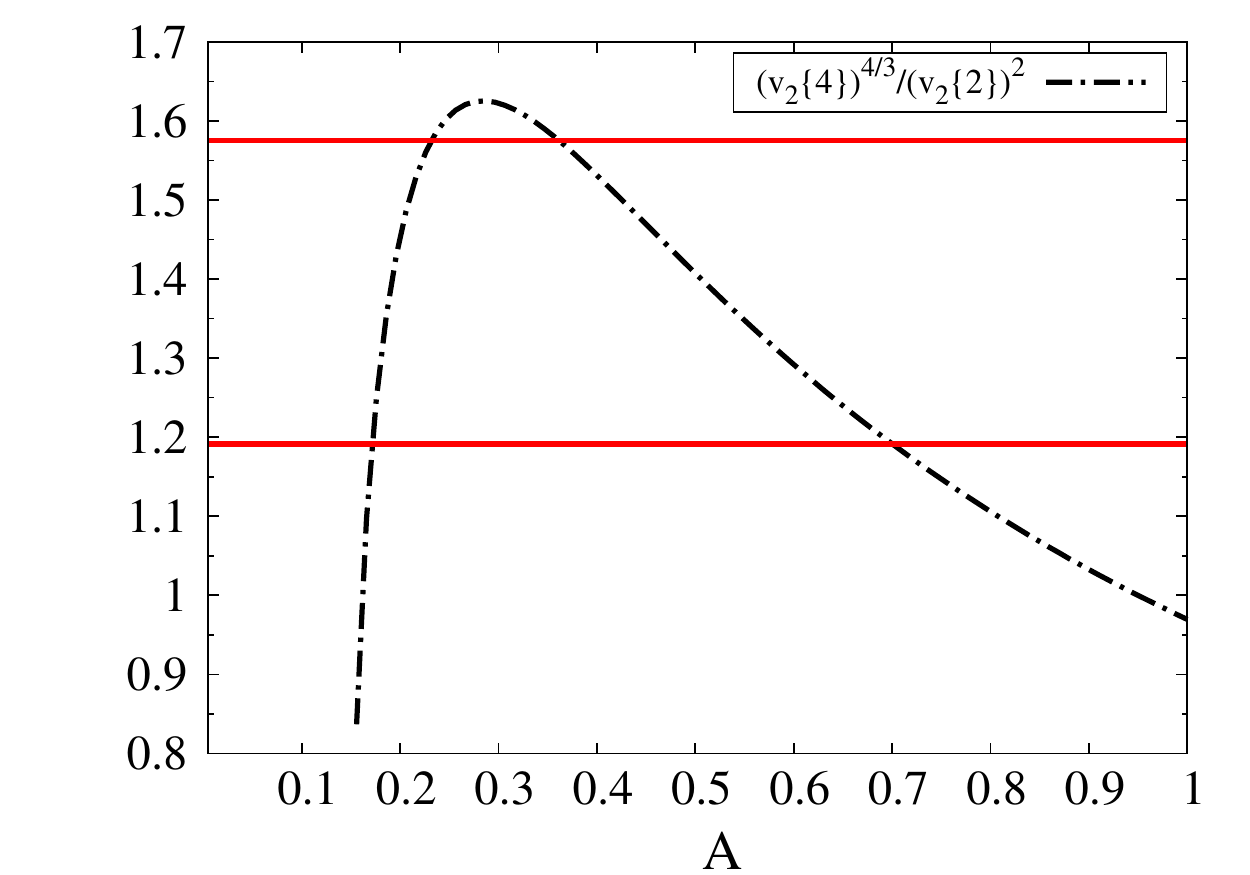}
\end{center}
\vspace*{-7mm}
\caption[a]{Left: Ratio of the full $v_{2}\{m\}$ ($m$ = 2, 3, 4) given by
  Eqs.~(\ref{Eq:c22},\ref{Eq:c24},\ref{eq:c23_end}) to the
  disconnected contribution corresponding to the first term of each of
  the equations, respectively.\\
  Right: Ratio $(v_{2}\{ 4\})^{4/3}/(v_{2}\{2\})^{2}$ as a function
  of the ${\bf E}$-field anisotropy amplitude, $\mathcal{A}$. The
  dash-dotted line corresponds to Eqs.~(\ref{Eq:c22},\ref{Eq:c24}).
  The straight horizontal lines are the values from the CMS data (see
  text for details).
}
\label{fig:full_to_disc_ratio}
\end{figure}
Note that these expressions do not include subleading corrections in
$N_c^{-2}$ which we defer to a future analysis. Strictly,
Eqs.~(\ref{Eq:c22b},\ref{Eq:c24b}) apply only for ${\cal A}={\cal
  O}(N_c^{-1})$ and ${\cal A}={\cal  O}(N_c^{-3/2})$,
respectively. Furthermore, the transverse momenta of {\em all}
particles are assumed to far exceed the saturation scale.

Depending on the value of $\mathcal{A}$ there are two different
regimes, see left panel of Fig.~\ref{fig:full_to_disc_ratio}: for
small values of $\mathcal{A}$ there are strong genuine
(non-factorizable) correlations and so the connected diagrams are
important; for large values of $\mathcal{A}$, however, the cumulants
approach the factorization limit where they are dominated by the fully
disconnected diagram and where genuine correlations are suppressed.

\begin{figure}[htb]
\begin{center}
\includegraphics[width=8cm]{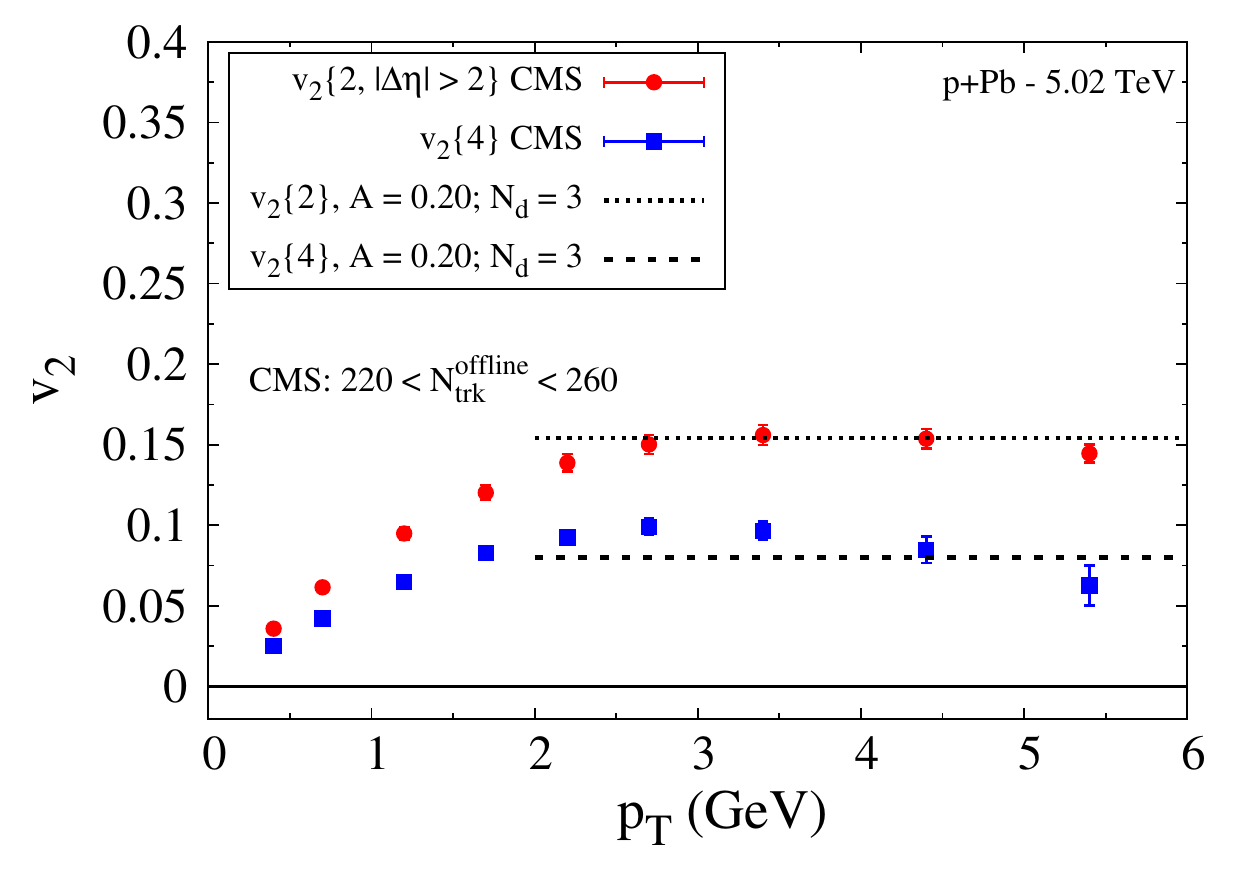}
\includegraphics[width=8cm]{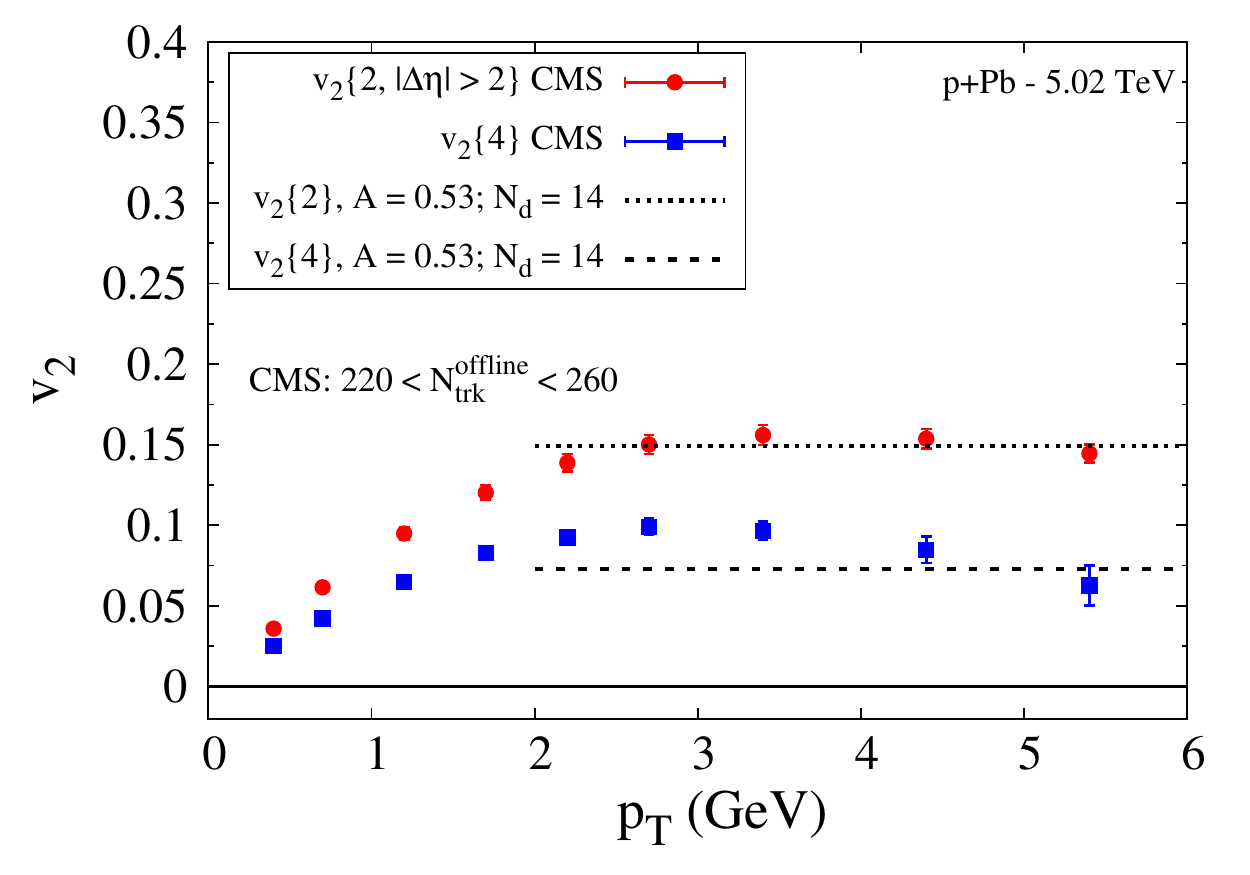}
\end{center}
\vspace*{-6mm}
\caption[a]{$v_2(p_T)$ from two- and four-particle cumulants for
  $\mathcal A=0.20$ (left) and $\mathcal A=0.53$ (right),
  respectively. In either case $N_D$ was fixed in order to reproduce the
  magnitudes of $v_{2}\{2\}$ and $v_{2}\{4\}$ simultaneously. The
  splitting between these two quantities is due to the contribution
  from connected diagrams. Data by CMS~\cite{pPb_CMS}.}
\label{fig:v22_v24_model_vs_data}
\end{figure}

In Fig.~\ref{fig:full_to_disc_ratio} (right panel) we plot the
$N_{D}$-independent ratio $(v_{2}\{ 4\})^{4/3}/(v_{2}\{2\})^{2}$ as a
function of $\mathcal{A}$. The dash-dotted line corresponds to
Eqs.~(\ref{Eq:c22},\,\ref{Eq:c24}) while the straight horizontal lines
represent that same ratio obtained from the two highest $p_T$ data
points for $v_{2}\{2\}$ and $v_{2}\{4\}$ shown in
Fig.~\ref{fig:v22_v24_model_vs_data}. As one can see the high-$p_T$
data allows two regimes of $\mathcal{A}$: one around $\mathcal{A}\sim
0.2$ and another for $0.35 \lesssim \mathcal{A} \lesssim 0.7$. From
Fig.~\ref{fig:full_to_disc_ratio} we see that the first solution is in
the regime where strong correlation effects are present while the
second one is close to the factorization limit.

The comparison of Eqs.~(\ref{Eq:c22},\ref{Eq:c24}) to the CMS
data in the high $p_{T}$ region is shown in
Fig.~\ref{fig:v22_v24_model_vs_data}.  The values of
$\mathcal{A}=0.20$ and $\mathcal{A}=0.53$ employed in the figure
correspond to the two possible solutions mentioned in the previous
paragraph. $N_{D}$ was fixed so as to reproduce the correct magnitudes
of $v_{2}\{2\}$ and $v_{2}\{4\}$.  Since both set of parameters, for
small and large $\mathcal{A}$, are able to describe the data with
comparable quality we must conclude that our analysis is not
sufficient to determine $\mathcal{A}$ and $N_D$ uniquely from the high
$p_T$ data alone.  If the data down to about $p_T=1$~GeV is included
in the analysis then the model would prefer smaller values
$\mathcal{A}\simeq0.2$~\cite{Dumitru:2014dra}.

\begin{figure}[htb]
\begin{center}
\includegraphics[width=8cm]{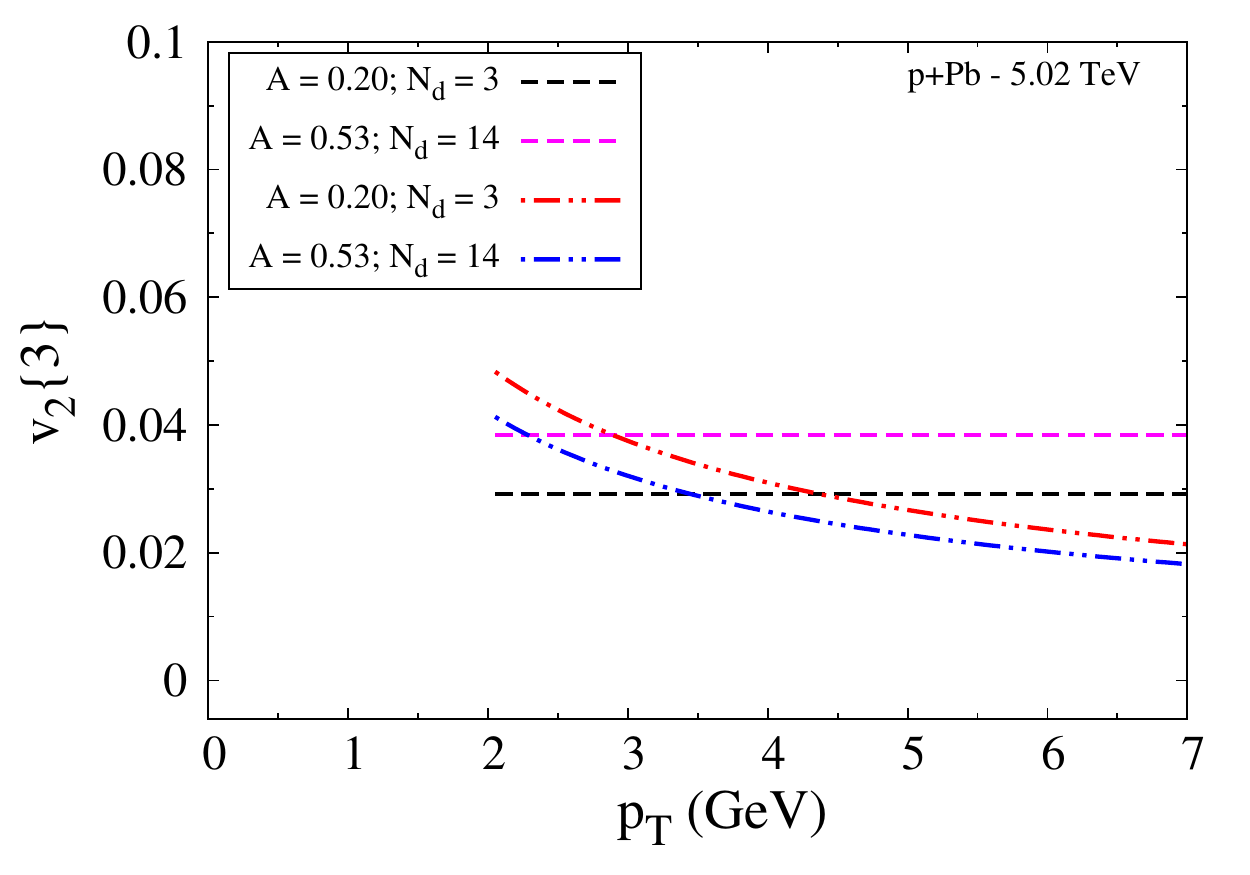}
\end{center}
\vspace*{-6mm}
\caption[a]{Prediction for $v_{2}\{3\}$ from
  Eqs.~(\ref{eq:c23_endb},\,\ref{c23_no_expansion_endb}). See text for
  details.}
\label{fig:v23_prediction}
\end{figure}

Figure~\ref{fig:v23_prediction} shows the predictions for $v_{2}\{3\}$
obtained from both expressions
\bea \label{eq:c23_endb}
c_{2}\{ 3 \} &=& (v_{2}\{3\})^{3} \sim
\frac{1}{N_{D}^{2}}\frac{1}{4}\frac{Q_{s}^{2}}{k_{3}^{2}}
\bigg( \frac{\mathcal{A}^{4}}{2} + \frac{\mathcal{A}^{2}}{2(N_{c}^2 - 1)} +
\frac{1}{16(N_{c}^{2} - 1)^{2}} \bigg)~, \\
c_{2}\{3\} &=& \frac{1}{N_{D}^{2}} \frac{A_4 \mathcal{A}^{2}}{8}
\label{c23_no_expansion_endb}
\eea
derived above, using the same values for $\mathcal A$ and $N_D$ as
deduced from $v_{2}\{2\}$ and $v_{2}\{4\}$. Recall
that~\eqref{eq:c23_endb} results from an expansion of the dipole
S-matrix to second order in $\tr (\rt\cdot{\bf E})^2$
while~\eqref{c23_no_expansion_endb} arises if the S-matrix exhibits a
$\cos(4\varphi)$ component already at order $r^2$. For
Fig.~\ref{fig:v23_prediction} we assumed that $A_4={\cal A}^2$ to avoid
introducing additional parameters.
The figure shows that despite the uncertainty in $\mathcal A$ and
$N_D$ that $v_{2}\{3\}$ does not vary too widely. We expect
$v_{2}\{3\}\approx 2-4\%$ for semi-hard transverse momenta.

\section{Summary and Outlook}

Our goal here was to provide a summary and overview of recent
ideas regarding anisotropic particle production at semi-hard
transverse momenta in high-energy collisions. The basic point is that
azimuthally anisotropic correlations should occur essentially due to
an anisotropic small-$x$ gluon distribution. 

The McLerran-Venugopalan model for the gluon distribution of dense
hadrons or nuclei integrates out the fast dynamics of the large-$x$
degrees of freedom and replaces them by ``frozen'' sources for
the small-$x$ semi-classical fields. We point out that each such
configuration exhibits azimuthal anisotropies with a finite transverse
correlation length, and that the angular structure of these
configurations is a {\em slow} variable, too.

As a consequence, before one averages over the random angular
structure of the source, the single-particle distribution due to
scattering of a projectile parton off such a target is
anisotropic. This gives rise to contributions to multi-particle
correlations from disconnected
diagrams~\cite{KovnerLublinsky,Dumitru:2014dra,Dumitru:2014yza}.  By
analogy to the BBGKY hierarchy the disconnected contributions dominate
the $m$-particle correlation functions in the limit of large
anisotropy ${\cal A}$ of the gluon distribution of the target. More
specifically, they have been shown~\cite{Dumitru:2014yza} to lead to a
sign flip of the four-particle ``elliptic'' cumulant $c_2\{4\}$.

On the other hand, the connected contributions to the cumulants from
small-$x$ dynamics exhibit a rather unexpected\footnote{From the point
  of view of conventional ``non-flow'' expectations.}  {\em coherence}
in that $c_2\{m\}$ depends weakly on the order $m$ of the cumulant,
for sufficiently large $m$~\cite{Skokov:2014tka}, quickly approaching
a constant as $1/m\to0$. Also, unlike conventional ``non-flow''
contributions, the connected diagrams from the CGC (coherent small-$x$
QCD dynamics) are long range in
rapidity~\cite{Dumitru:2008wn,pAridge}.

Much work is still needed before we might claim to understand
the data. From the point of view of phenomenology one should, for
example, compute subleading in $N_c^{-2}$ and $Q_s^2/p_T^2$
corrections to the cumulants in order to improve the analysis
performed in sec.~\ref{sec:FitsResults}. It would be very interesting,
too, to resum the time evolution of the gluon fields in the forward
light cone~\cite{SchlichtingIS2014} in order to understand over what
range of $p_T$ final-state interactions are important. Also, this
approach could clarify the time scale over which odd-index
cumulants like $c_1\{m\}$ and $c_3\{m\}$ develop. As a last point, let
us mention that the effect of the multiplicity bias employed in
experiments on the anisotropy of the gluon field configurations is
poorly understood at present.

An interesting theoretical issue is the relationship of the azimuthal
cumulants to the gluon distributions introduced (via specific
operator-level relations) in TMD factorization~\cite{TMD}. Within the
MV model at least, it has been shown by explicit computation that
${\cal A}(r)$, which corresponds to $v_2\{1\}(p_T)$ at high $p_T$,
coincides with the distribution of linearly polarized gluons
$h_1^{\perp g}(r)$~\cite{TMD2}. It remains to be seen if this
relation still applies after resummation of small-$x$ quantum
corrections. Azimuthal correlations in high-energy p+p and p+A
collisions could provide much insight into non-trivial QCD
dynamics.

\section*{Acknowledgements}
We thank Yu.~Kovchegov, A.~Kovner, T.~Lappi, M.~Lublinsky, L.~McLerran,
R.~Venugopalan, and S.~Schlichting for many
useful discussions.  A.D.\ gratefully acknowledges support by the DOE
Office of Nuclear Physics through Grant No.\ DE-FG02-09ER41620 and
from The City University of New York through the PSC-CUNY Research
Award Program, grant 66514-0044. A.V.G. gratefully acknowledges the
Brazilian Funding Agency FAPESP for financial support (contract:
2013/23848-5).


\end{document}